\documentclass[usenatbib,referee]{mn2e}
\usepackage{amsmath}
\usepackage{graphicx}
\usepackage{epsfig}
\usepackage{txfonts}
\usepackage{color}
\usepackage{natbib}

\title[The hot subdwarfs from the surviving companions of SNe Ia]
{Hot subdwarfs from the surviving companions of the white dwarf +
main-sequence channel of Type Ia supernovae}
\author[Meng \& Luo]
{Xiang-Cun Meng$^{\rm 1,2,3}$\thanks{E-mail:xiangcunmeng@ynao.ac.cn}, Yang-Ping Luo$^{\rm 4}$\\
$^1$Yunnan Observatories, Chinese Academy of Sciences, 650216 Kunming, PR China\\
$^2$Key Laboratory for the Structure and Evolution of Celestial
Objects, Chinese Academy of Sciences, 650216 Kunming, PR
China\\
$^3$Center for Astronomical Mega-Science, Chinese Academy of
Sciences, 20A Datun Road, Chaoyang District, Beijing, 100012, P.
R. China\\
$^{\rm 4}$Department of Astronomy, China West Normal University,
Nanchong, 637002, PR China}

\begin{document}
\date{}
\pagerange{\pageref{firstpage}--\pageref{lastpage}} \pubyear{2018}
\maketitle

\label{firstpage}

\begin{abstract}\label{abstract}
Some surviving companions of type Ia supernovae (SNe Ia) from the
white-dwarf + main-sequence (WD + MS) channel may evolve to hot
subdwarfs. In this paper, we preformed stellar evolution
calculations for surviving companions of close WD + MS systems in
the spin-up/spin-down model and the canonical non-rotating model
to map out the initial parameter spaces in the orbital period -
secondary mass plane in which the surviving companions can evolve
to hot subdwarfs. Based on these results, we carried out a series
of binary population synthesis calculation to obtain the Galactic
birth rate of the hot subdwarfs from the WD + MS channel, which is
$2.3-6\times10^{\rm -4}\,{\rm yr}^{\rm -1}$ for the
spin-up/spin-down model and $0.7-3\times10^{\rm -4}\,{\rm yr}^{\rm
-1}$ for the canonical non-rotating model. We also show the
distributions of some integral properties of the hot subdwarfs,
e.g. the mass and the space velocity, for different models. In
addition, comparing our results with the observations of the
intermediate helium-rich (iHe-rich) hot subdwarfs, the hot
subdwarfs from the WD + MS channel may explain some observational
features of the iHe-rich hot subdwarfs, especially for those from
the spin-up/spin-down model. Although we expect that the SN Ia
channel can only contribute a small fraction of the iHe-rich hot
subdwarf population, some of these may help to explain cases with
unusual kinematics.
\end{abstract}

\begin{keywords}
stars: supernovae: general - stars: chemically peculiar - stars:
kinematics and dynamics - subdwarfs
\end{keywords}

\section{INTRODUCTION} \label{sect:1}
As one of the best distance indicators because of their remarkable
uniformity and high luminosity, Type Ia supernovae (SNe Ia) were
used to measure the cosmological parameters, which led to the
discovery of an accelerating expansion of the Universe and a
mysterious dark energy (\citealt{RIE98}; \citealt{PER99}). SNe Ia
are used to be cosmological probes for testing the
equation-of-state of the dark energy (EOSDE) and its evolution
with time (\citealt{HOWELL11}; \citealt{SULLIVAN11}). SNe Ia are
also very important to understand galactic chemical evolution
owing to their major contribution of iron to their host galaxies
(\citealt{GREGGIO83}; \citealt{MATTEUCCI86}).

Although SNe Ia are so important in modern astrophysical fields,
some basic problems with SNe Ia are still under debate, especially
on their progenitor systems (\citealt{HN00}; \citealt{LEI00}). It
is now widely believed that a SN Ia is derived from a binary
system with at least one carbon oxygen white dwarf (CO WD,
\citealt{HN00}; \citealt{NUGENT11}). The CO WD accretes material
from its companion to increase its mass until its mass reaches a
maximum stable value to trigger a thermonuclear runaway in the WD.
The released nuclear energy unbinds the WD and then a SN Ia is
produced (\citealt{BRA04}; \citealt{HILLEBRANDT13}). The companion
of the CO WD may be a main-sequence or a slightly evolved star
(WD+MS), a red giant star (WD+RG) or a helium  star (WD + He
star). This is the single degenerate (SD) model (\citealt{WI73};
\citealt{NTY84}). Or it may be another CO WD involving the merging
of two CO WDs. This is the double degenerate (DD) model
(\citealt{IT84}; \citealt{WEB84}). At present, both models obtain
some support on both observational and theoretical sides
(\citealt{RUIZLAPUENTE19}; \citealt{JHA19}).

A basic way to distinguish between the SD and the DD models is to
search for surviving companions in supernova remnants (SNRs)
because the SD model predicts the existence of surviving
companions in SNRs but not from the DD model. It shows the power
of the method for the discovery of some potential surviving
companions in some SNRs (\citealt{RUIZLAPUENTE04};
\citealt{LIC17}). However, many teams also reports that they did
not find any candidate of the surviving companions in other SNRs.
This seems to favour the DD model (\citealt{GONZALEZ12};
\citealt{SCHAEFER12}; \citealt{RUIZLAPUENTE17};
\citealt{KEERZENDORF17}). A possible solution for this
embarrassment of the SD model is the so-called spin-up/spin-down
model. The WDs may experience a long spin-down phase before
supernova explosion and then the companions become too dim to be
detected (\citealt{JUSTHAM11}; \citealt{DISTEFANO12}). Before
supernova explosion, some companions are proposed to become
low-mass helium WDs (\citealt{JUSTHAM11}; \citealt{DISTEFANO12}),
but such a suggestion has not been confirmed by searching for the
surviving WD companion in SN 1006 (\citealt{KEERZENDORF17}). One
possible reason is the huge uncertainty of the spin-down timescale
(\citealt{DISTEFANO11}; \citealt{MENGXC13}). Recently,
\citet{MENGXC19a} noticed that some surviving companions from the
WD + MS channel are hot subdwarfs at the supernova moment (see
also the discussion by \citealt{JUSTHAM11}), and then they
suggested that the surviving companions of SN 1006 and Kepler's
supernova could be hot subdwarf B (sdB) stars\footnote{The
surviving companions of SNe Ia from the WD + He star channel are
single hyper-velocity hot subdwarfs (\citealt{JUSTHAM09};
\citealt{WANGB09}), but because of mass transfer and tight orbit
before supernova explosion, such hot subdwarfs are more likely to
be extremely helium-rich ones with rapidly rotational velocity, as
observed in US 708 and J2050 (\citealt{GEIER15};
\citealt{ZIEGERER17}). In this paper, we do not want to study
those high-velocity runaway hot subdwarfs from SNe Ia with
helium-rich donors in short-period orbits (\citealt{WANGB09}).}.
Hot subdwarfs are generally helium-core-burning stars with
extremely thin hydrogen-rich envelopes
(\citealt{HEBER09,HEBER16}).

In addition, a surviving companion may not always associate with a
SNR because its life is much longer than the typical life time of
a SNR (a few $10^{\rm 4}\,$yr, \citealt{SARBADHICARY17}). For
example, \citet{PIETRUKOWICZ17} discovered a new kind pulsator,
the blue large-amplitude pulsators (BLAPs) which are probably
single stars. Based on the common envelope wind (CEW) model
developed by \citet{MENGXC17a}, \citet{MENGXC20} found that all
the properties of BLAPs can be reproduced by the surviving
companions of SNe Ia from the WD + MS channel, and they are also
core-helium-burning stars, similar to hot subdwarfs. Moreover,
\citet{MENGXC20} also noticed that some surviving companions of
SNe Ia from the WD + MS channel may evolve to hot subdwarfs if
their hydrogen-rich envelope is thin enough at the moment of the
supernova explosion. The surface helium abundance and the
effective temperature of the hot subdwarfs from the WD + MS
channel are consistent with the intermediate He-rich (iHe-rich)
stars. However, \citet{MENGXC20} did not consider the effect of
the spin-up/spin-down model. In particular, they did not show the
initial parameter space, and then the birth rate of the hot
subdwarfs from the WD + MS channel. Following \citet{MENGXC19a}
and \citet{MENGXC20}, we want to study the production of the hot
subdwarfs from the WD + MS channel in details, especially their
birth rate. In addition, we also try to constrain the contribution
of the hot subdwarfs from the WD + MS channel to the iHe-rich
population by a large sample from the survey of the Large Sky Area
Multi-Object Fiber Spectroscopic Telescope (LAMOST).

In Sec.~\ref{sect:2}, we describe our methods and then present the
calculation results in Sec.~\ref{sect:3} and Sec.~\ref{sect:4}. In
Sec.~\ref{sect:5}, we summarize the properties of the hot
subdwarfs from the WD + MS channel and compare our results with
observations. We discuss the uncertainties of our results and the
other possible contributors to the iHe-rich hot subdwarf stars in
Sec.~\ref{sect:6} and summarize our main conclusions in
Sec.~\ref{sect:7}.

\section{METHODS}\label{sect:2}
Following \citet{MENGXC20}, we use the CEW model to calculate the
binary evolution of the WD + MS systems, where the mass transfer
from the companions to the WDs may begin when the companions are
on the MS or in the Hertzsprung gap (HG). The basic prescriptions
of binary evolution for the WD + MS systems are the same to those
used by \citet{MENGXC17a}, and so we do not give the repetitious
details here. As far as the properties of the surviving companion
after the supernova explosion are concerned, the difference for
most cases between the CEW and the optically thick wind (OTW)
models is not very significant (\citealt{HAC96}). In particular,
the distributions of the initial binary parameters, i.e. initial
WD masses, initial companion mass and initial orbital period,
between the CEW and the OTW models are also similar because of
similar initial parameter space for SNe Ia. However, as shown by
\citet{MENGXC17a}, some systems that cannot produce SNe Ia in the
OTW model do so in the CEW model.

In theory, the accreting WD may spin up by gaining the angular
momentum of the accreted materials and a rapidly rotating WD may
even increase its mass to exceed 1.378 $M_{\odot}$ rather than
explode immediately (\citealt{YOON04,YOON05}). The rotating
super-Chandrasekhar WDs will experience a spin-down phase before
they explode as SNe Ia (\citealt{JUSTHAM11};
\citealt{DISTEFANO12}; \citealt{HACHISU12}). The spin-down phase
is also required by the observations of some SNe Ia
(\citealt{SOKER18,SOKER19}). However, at present, there are still
many uncertainties on the spin-up/spin-down model, e.g. the
spin-down timescale although a few $10^{\rm 6}$ yr is favoured
(\citealt{MENGXC18a}; \citealt{SOKER19}), the exact time of the
onset of the spin-down phase (\citealt{MENGXC13}) and WD growth
pattern after $M_{\rm WD}=1.378~M_{\odot}$. So, following
\citet{MENGXC19a}, we define a pseudo spin-down timescale,
$\tau_{\rm sp}$, that is the time interval from when $M_{\rm
WD}=1.378~M_{\odot}$ to the time of supernova explosion. Here, we
set $\tau_{\rm sp}=10^{\rm 7}$ yr for the spin-up/spin-down
model\footnote{Based on the definition of the pseudo spin-down
timescale here, a real spin-down timescale for the cases
considered here is shorter than the pseudo spin-down timescale of
$10^{\rm 7}$ yr (\citealt{MENGXC19a}).}. During the pseudo
spin-down phase, we simply assume the same WD growth pattern as
for $M_{\rm WD}<1.378~M_{\odot}$ (\citealt{MENGXC17a};
\citealt{MENGXC19a}). As a comparison, we also consider a case
with $\tau_{\rm sp}=0$ , which corresponds to the canonical
non-rotating model as in \citet{MENGXC20}. Then, in the following
part of the paper, the cases with $\tau_{\rm sp}=10^{\rm 7}$ yr
and $\tau_{\rm sp}=0$ yr refer to those from the spin-up/spin-down
model and the canonical non-rotating model, respectively.

After the supernovae, we continue to evolve the surviving
companions as single stars with a Reimers's wind with $\eta=0.25$
(\citealt{REIMERS75}; \citealt{FUSI76}). To study the birth rate
of the hot subdwarfs from the WD + MS channel, we calculated a
dense model grid with different initial WD masses, different
initial secondary masses and different initial orbital periods to
check the parameter space in which the surviving companions can
evolve to hot subdwarfs. The initial masses of the companions,
$M_{\rm 2}^{\rm i}$, range from 2.2 to 4.0 $M_{\odot}$ with a step
of 0.1 $M_{\odot}$. Because hybrid carbon-oxygen-neon WDs (CONe
WDs) could produce some special SNe Ia, e.g. SN 2002cx-like or SN
Ia-CSM objects, we assume that a WD with a mass as massive as 1.3
$M_{\odot}$ can lead to a SN Ia (\citealt{CHENMC14};
\citealt{MENGXC14,MENGXC18a}). Then, the initial masses of the
WDs, $M_{\rm WD}^{\rm i}$, range from 0.8 to 1.30 $M_{\odot}$ with
a step of 0.1 $M_{\odot}$. The initial orbital periods of binary
systems, $P^{\rm i}$, from 1 day, at which the companions fill
their Roche lobes at MS, to about $15$ day, at which the
companions fill their Roche lobes at the end of the HG, with a
step of 0.1 in $\log_{\rm 10}(P^{\rm i}{\rm /day})$. We so obtain
the parameter space for the hot subdwarfs from the WD + MS
channel.

Based on this parameter space, we made a series of binary
population synthesis (BPS) calculations using the rapid binary
evolution code developed by \citet{HUR00} and \citet{HUR02}. If a
binary system in the simulations evolves to the WD+MS stage and is
at the onset of Roche-lobe overflow (RLOF) and is located in the
($\log P^{\rm i}$, $M_{\rm 2}^{\rm i}$) plane for a SN Ia whose
surviving companion can evolve to a hot subdwarf, we assume that
the supernova explosion leads to the formation of a hot subdwarf.
We followed the evolution of $10^{\rm 8}$ sample binaries, where
the primordial binary samples are generated in a Monte Carlo way.
For the Monte Carlo simulations, we use (1) a constant star
formation rate of 5 $M_{\odot}~{\rm yr^{\rm -1}}$
(\citealt{WK04}), or a single starburst of $10^{\rm 11}$
$M_{\odot}$, (2) the initial mass function (IMF) of \citet{MS79},
(3) a uniform mass-ratio distribution, (4) a uniform distribution
of separations in $\log a$ for binaries, where $a$ is the orbital
separation, (5) circular orbits for all binaries, and (6) For BPS
simulations, the common-envelope ejection efficiency, $\alpha_{\rm
CE}$, is the key parameter to affect the birth rate of hot
subdwarf stars. Following \citet{MENGXC17a}, we take $\alpha_{\rm
CE}=1.0$ or $\alpha_{\rm CE}=3.0$ (see \citealt{MENG09} and
\citealt{MENGXC17a} for details).

\section{BINARY EVOLUTIONARY RESULTS}\label{sect:3}
Whether or not the surviving companion of a SN Ia may evolve to a
hot subdwarf depends on the initial parameters of a WD + MS
system. In this section, we present how theses parameters affect
the evolution of surviving companions, and summarize our stellar
evolutionary outcomes.

\begin{figure}
\centerline{\includegraphics[angle=270,scale=.35]{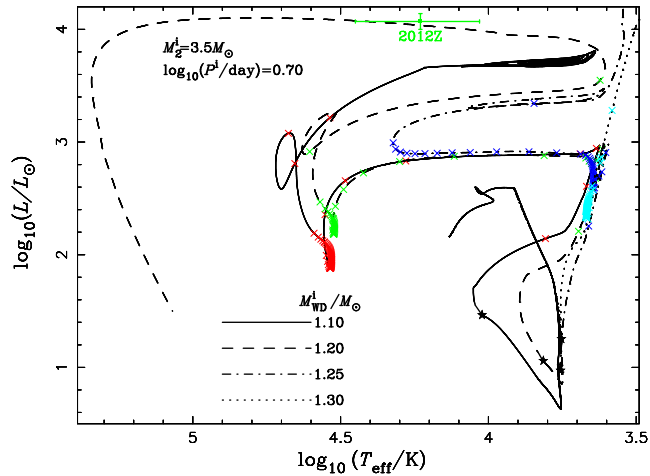}}
\caption{The evolution of the companions in the HR diagram from
the binary systems with different initial WD masses and $\tau_{\rm
sp}=0$, where the initial companion mass and the initial orbital
period are [$M_{\rm 2}^{\rm i}, \log_{\rm 10}(P^{\rm i}{\rm
/day})]$=[$3.5\,M_{\odot}$, 0.7]. The black stars show the
positions where supernova explosions are assumed. The age interval
between adjacent crosses in every line is $10^{\rm 6}$ yr. The
green cross with label `2012Z' presents the suggested companion of
SN 2012Z (\citealt{MCCULLY14}).}\label{hrdwd}
\end{figure}

\subsection{Dependence on the initial binary parameters }\label{sect:3.1}
\subsubsection{Dependence on the initial WD mass }\label{sect:3.2.1}
In Fig.~\ref{hrdwd}, we show the evolution of the companions in
the  Hertzsprung--Russell (HR) diagram from the systems with
different initial WD masses and $\tau_{\rm sp}=0$, where the
initial companion mass and the initial orbital period are [$M_{\rm
2}^{\rm i}, \log_{\rm 10}(P^{\rm i}{\rm
/day})]$=[$3.5\,M_{\odot}$, 0.7]. For these cases, the companions
fill their Roche lobes in the HG and then mass transfer begins. In
Fig.~\ref{hrdwd}, the surviving companions from the systems with
$M_{\rm WD}^{\rm i}=1.1$ $M_{\odot}$ and $M_{\rm WD}^{\rm i}=1.2$
$M_{\odot}$ can become hot subdwarfs. However, for the case with
$M_{\rm WD}^{\rm i}=1.3$ $M_{\odot}$, the following evolution of
the surviving companion, after supernova explosion, is quite
similar to that of a single star. The star consecutively
experiences the RG, horizontal branch (HB) and AGB phases.
Interestingly, there is a transitional case when $M_{\rm WD}^{\rm
i}=1.25$ $M_{\odot}$. For this case, when the star evolves into
the HB phase, its hydrogen-rich envelope is not very thick. With
the consumption of the envelope by shell hydrogen burning, the
envelope becomes thinner and thinner, and then the effective
temperature may become as high as 20000 K. Such different
evolutions of the companions after supernova explosion are derived
from the different hydrogen-rich envelope masses at the moment of
the supernova explosion, i.e. the thicker the envelope, the more
similar is the evolution of the surviving companion to a single
star. For a system with a given initial companion mass and a given
initial orbital period, a less massive initial WD means more
accreted material from its companion before it explodes as a SN Ia
and a less massive companion with a thinner envelope.

Once the helium in the centre of the surviving companion is
exhausted, some hot subdwarfs evolve to a helium RG phase and
finally become WDs, while the others become WDs directly, without
a helium RG phase\footnote{As shown in Fig.~\ref{hrdwd}, in our
calculations, some companions experience a series of hydrogen
flashes after their helium RG phases. For such cases, we stop our
calculation after several hydrogen flashes because this does not
affect our discussion on hot subdwarfs.}. Such different
evolutions are mainly caused by the different helium envelope
masses when the helium is exhausted in the centre. Generally, if
the helium envelope is more massive than 0.3 $M_{\odot}$, the hot
subdwarfs experience a RG phase (\citealt{JUSTHAM11b}). In
addition, in Fig.~\ref{hrdwd}, we also plot the proposed companion
to SN 2012Z (\citealt{MCCULLY14}), whose position in the HR
diagram is consistent with a helium RG star as discussed by
\citet{WANGB14b} and \citet{LIUZW15}.

\begin{figure}
\centerline{\includegraphics[angle=270,scale=.35]{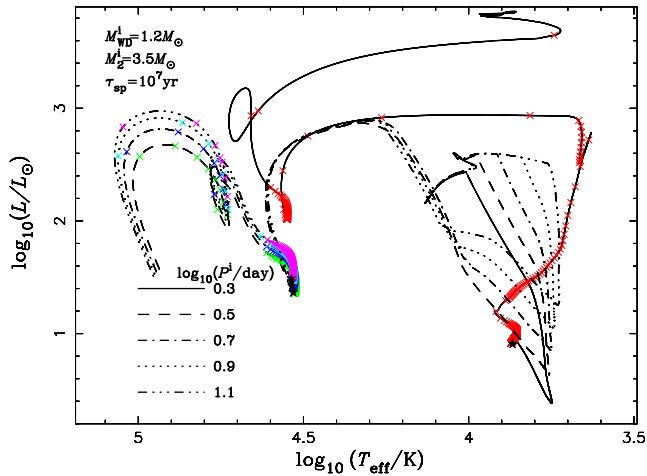}}
\caption{The evolution of the companions in the HR diagram from
the systems with different initial orbital periods and $\tau_{\rm
sp}=10^{\rm 7}$ yr, where the initial WD mass and the initial
secondary mass are 1.2 $M_{\odot}$ and 3.5 $M_{\odot}$,
respectively. The black stars show the positions where supernova
explosions are assumed. The age interval between adjacent crosses
in every line is $10^{\rm 6}$ yr. }\label{hrdpersp7}
\end{figure}

\subsubsection{Dependence on the initial orbital period }\label{sect:3.1.2}
In Fig.~\ref{hrdpersp7}, we show the evolution of the companions
in the HR diagram from the systems with different initial orbital
periods and $\tau_{\rm sp}=10^{\rm 7}$ yr, where the initial WD
mass and the initial secondary mass are 1.2 $M_{\odot}$ and 3.5
$M_{\odot}$, respectively. Fig.~\ref{hrdpersp7} shows that, for
the cases in which mass transfer begins when the companions are in
the HG, the companions are hot subdwarfs at the moment of
supernova explosion because of a long spin-down timescale. After
the exhaustion of the central helium, the hot subdwarfs evolve to
the WD branch directly owing to their low mass.

However, for the system with $\log_{\rm 10}(P^{\rm i}{\rm
/d})=0.3$, mass transfer begins when the companion is on the MS,
and then the companion is still a MS star at the moment of the
supernova explosion. However, the surviving companion still can
evolve to the hot subdwarf stage after the MS, RG and a short-life
HB phase. The luminosity of the hot subdwarf is significantly
higher than those of the other cases. After helium is exhausted in
the centre of the hot subdwarf, it experiences a helium RG phase
before it enters the WD branch for a relatively large mass.
Therefore, the surviving companion consecutively experiences the
MS, HG, RG, HB, sdB, helium RG and WD phase within about 350 Myr.

\begin{figure}
\centerline{\includegraphics[angle=270,scale=.35]{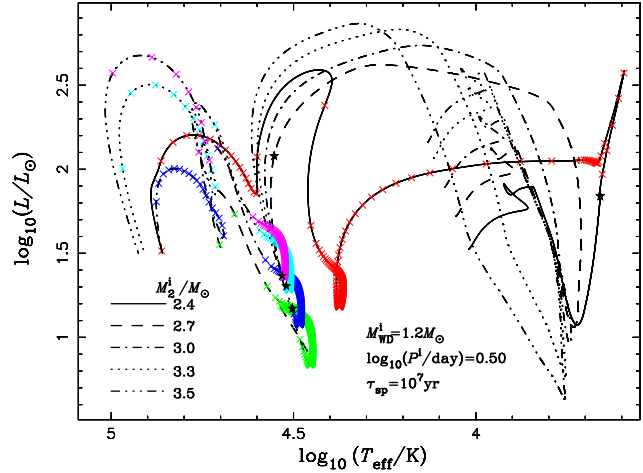}}
\caption{The evolution of the companions in the HR diagram of the
systems with different initial companion masses and $\tau_{\rm
sp}=10^{\rm 7}$ yr, where the initial WD mass and the initial
orbital period are [$M_{\rm WD}^{\rm i}, \log_{\rm 10}(P^{\rm
i}{\rm /day})]$=[$1.2\,M_{\odot}$, 0.5]. The black stars show the
positions where supernova explosions are assumed. The age interval
between adjacent crosses in every line is $10^{\rm 6}$ yr.
}\label{hrdm2sp7}
\end{figure}

\subsubsection{Dependence on the initial companion mass }\label{sect:3.1.3}
Similarly, in Fig.~\ref{hrdm2sp7}, we show the evolutionary tracks
of the companions in the HR diagram, where the initial binary
systems with [$M_{\rm WD}^{\rm i}, \log_{\rm 10}(P^{\rm i}{\rm
/day})]$ = [$1.2\,M_{\odot}$, 0.5] have different initial
companion masses. The basic evolutionary tracks are similar to
those shown in Fig.~\ref{hrdpersp7}. For the system with $M_{\rm
2}^{\rm i}=2.4$ $M_{\odot}$, the supernova occurs when the
companion is a RG star (see also \citealt{MENGXC19a}), and then
there is still a relatively thick hydrogen-rich envelope at the
moment of the supernova explosion. This is main reason why it has
a lower effective temperature.

\begin{figure}
\begin{minipage}[t]{0.5\textwidth}
    \centering
\includegraphics[angle=270,scale=.35]{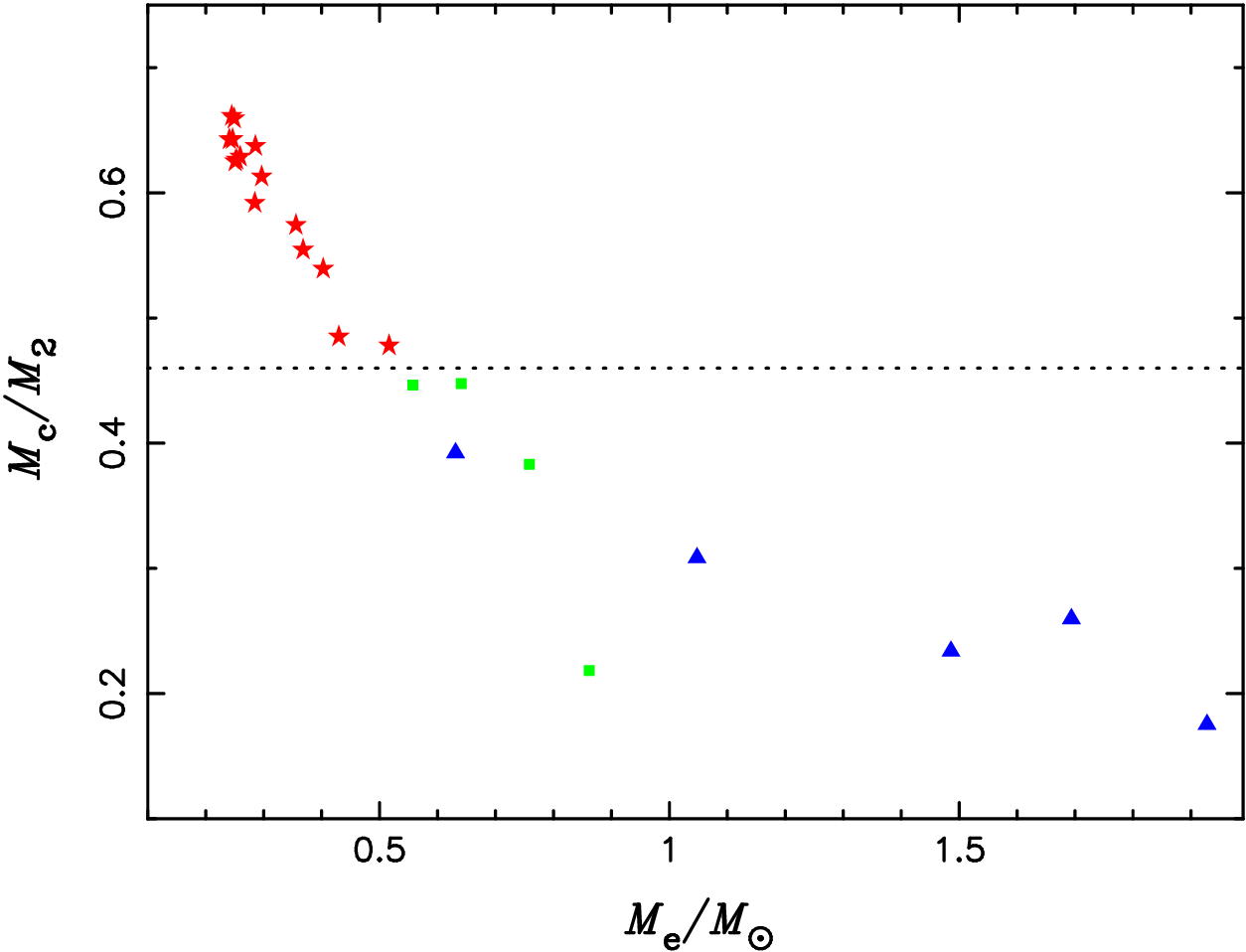}
\end{minipage}
\begin{minipage}[t]{0.5\textwidth}
    \centering
\includegraphics[angle=270,scale=.35]{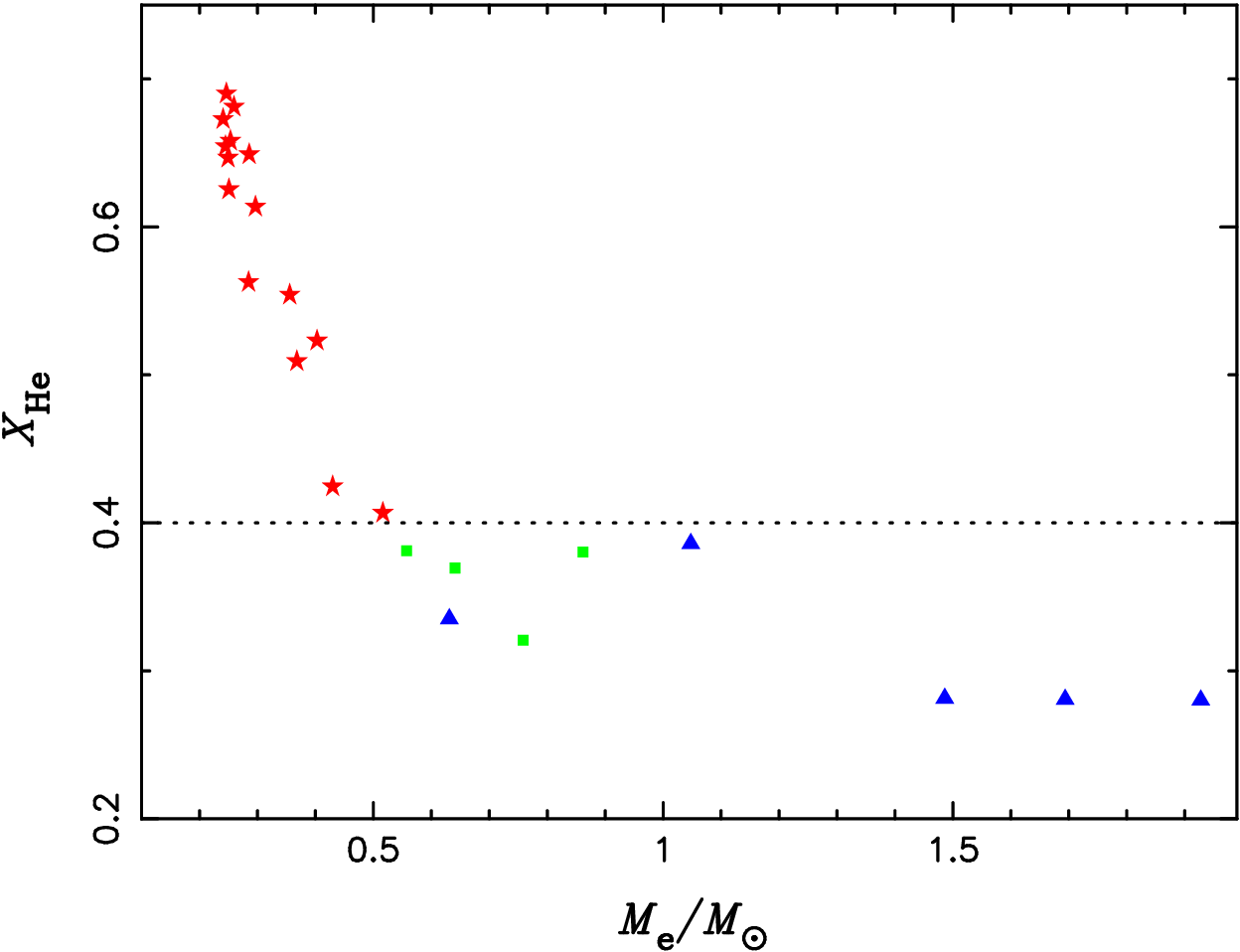}
\end{minipage}
\caption{Left: the ratio of the core mass to the companion mass
vs. the envelope mass at the moment of supernova explosion. Right:
the surface helium abundance vs. the envelope mass at the moment
of supernova explosion. The red stars represent those for which
the surviving companion can evolve to the hot subdwarf phase,
while the blue triangles represent those for which the surviving
companion cannot evolve to the hot subdwarf phase. The green
squares show the transitional cases similar to those shown in
Fig.~\ref{hrdwd}.}\label{memc}
\end{figure}


\subsection{The properties of the companions at supernova moment}\label{sect:3.2}
According to above discussions, whether a surviving companion may
or not become a hot subdwarf is heavily dependent on the envelope
mass and the core mass of the companion at the moment of the
supernova explosion, i.e. the thinner the envelope of the
companion, the more likely is the star to evolve to the hot
subdwarf phase. In Fig.~\ref{memc}, we show the correlation
between the core to companion mass ratio ($M_{\rm c}/M_{\rm 2}$)
and the envelope mass ($M_{\rm e}$) at the moment of the supernova
explosion (left panel), and the surface helium abundance of the
companion vs. the envelope mass (right panel). All the systems
discussed in Sec.~\ref{sect:3.1} are included in Fig.~\ref{memc}.
The figure clearly shows that the more massive the envelope, the
lower the ratio of the core mass to the companion mass, and the
lower the surface helium abundance. There are boundaries for the
stars that can evolve to hot subdwarfs, $M_{\rm c}/M_{\rm
2}>0.46$, $M_{\rm e}<0.52$ $M_{\odot}$, and $X_{\rm He}>0.4$.
Moreover, because the lifetimes of the transitional cases (green
squares in Fig.~\ref{memc}) with a high effective temperature is
short, we exclude them from the hot subdwarfs in our remaining
discussion. Nonetheless, \citet{MENGXC20} verified that the
properties of the BLAPs are consistent with core-helium-burning
stars, and then in this paper, we do not discriminate the BLAPs
from the hot subdwarfs.

\begin{figure}
\centerline{\includegraphics[angle=270,scale=.35]{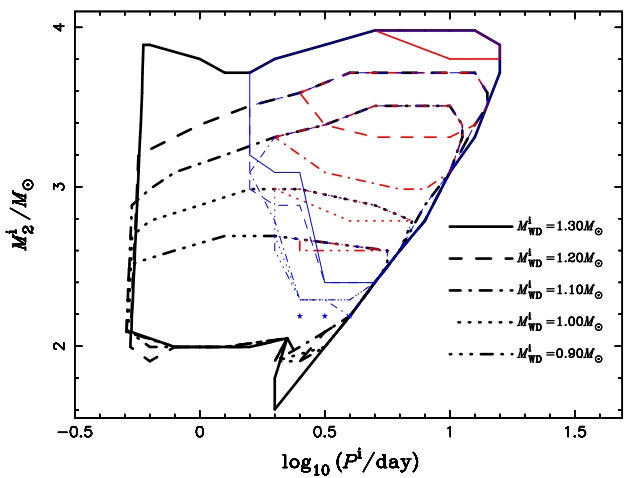}}
\caption{Parameter space leading to hot subdwarf stars in the
initial orbital period - secondary mass plane for different
initial WD mass. Red lines show the cases with $\tau_{\rm sp}=0$
yr, while the blue lines present the cases with $\tau_{\rm
sp}=10^{\rm 7}$ yr. For the cases with $M_{\rm WD}^{\rm i}=0.8$
$M_{\odot}$ and $\tau_{\rm sp}=10^{\rm 7}$ yr, only three models
can produce the hot subdwarf surviving companions, as shown by the
blue stars. The black lines show the contours leading to SNe Ia,
where the data are from \citet{MENGXC17a,MENGXC18a}.}\label{gsdb}
\end{figure}

\subsection{Parameter space for hot subdwarfs}\label{sect:3.3}
All the surviving companions from the WD + He star channel show
the properties of hot subdwarf or helium RG stars
(\citealt{WANGB14b}; \citealt{LIUZW15}), but not all the surviving
companions from the WD + MS channel may evolve to the hot
subdwarfs. In Fig.~\ref{gsdb}, we show the parameter space leading
to the hot subdwarfs in the initial orbital period -- secondary
mass ($\log P^{\rm i}-M_{\rm 2}^{\rm i}$) plane with $\tau_{\rm
sp}=0$ and $\tau_{\rm sp}=10^{\rm 7}$ yr. The figure shows that no
matter how long is the spin-down timescale, some binary systems
always produce hot subdwarfs. Compared with the initial parameter
space for SNe Ia, the parameter space for the hot subdwarfs
locates at the upper--right region in the $\log P^{\rm i}-M_{\rm
2}^{\rm i}$ plane. The parameter space leading to hot subdwarfs
for the spin-up/spin-down model is larger than that for the
canonical non-rotating model. This is easy to understand because a
longer spin-down timescale means a thinner envelope of the
companion at the moment of supernova explosion and then the
companion is more likely to evolve to the hot subdwarf phase. So,
we expect that the spin-up/spin-down model leads to a higher birth
rate of the hot subdwarfs than the canonical non-rotating model.
In addition, for the models with $\tau_{\rm sp}=10^{\rm 7}\,{\rm
yr}$, many systems for which mass transfer begins at the end of
the MS may produce hot subdwarfs, while for the models with
$\tau_{\rm sp}=0$ yr, only the system for which mass transfer
begins at the HG stage can produce hot subdwarfs. In particular,
if $M_{\rm WD}^{\rm i}<0.8$ $M_{\odot}$, no surviving companion
can evolve to the hot subdwarf phase. This does not depend on the
spin-down timescale. Therefore, the majority of SNe Ia cannot
produce the hot subdwarfs because the distribution of the initial
WD mass for SNe Ia peaks at $M_{\rm WD}^{\rm i}\approx0.78$
$M_{\odot}$ (\citealt{MENG09}).

\begin{figure}
\centerline{\includegraphics[angle=270,scale=.35]{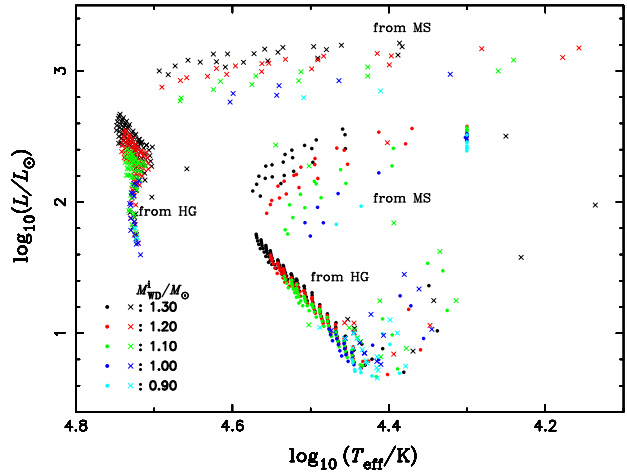}}
\caption{The outcomes of the evolution of the surviving companions
for different initial WD masses in the HR diagram, where
$\tau_{\rm sp}=10^{\rm 7}$ yr. The dots represent the times at
which the surviving companions have the lowest luminosity during
the hot subdwarf phase, while the crosses represent the times at
which the helium is exhausted in the centres of the surviving
companions. The symbols `MS' and `HG' in the figure mean that, at
the onset of the mass transfer, the companions are on the MS or in
the HG, respectively}\label{hrdspwd}
\end{figure}

\subsection{The outcomes of the evolution of the surviving companions}\label{sect:3.4}
In Fig.~\ref{hrdspwd}, we show the outcomes of the evolution of
the surviving companions for different initial WD masses in the HR
diagram, in which the dots represent the times for which the
surviving companions have the lowest luminosity during the hot
subdwarf phase, while the crosses represent the times for which
helium is exhausted in the centres of the surviving companions.
Because the initial parameter space for $\tau_{\rm sp}=10^{\rm 7}$
yr covers that for $\tau_{\rm sp}=0$ yr, we here only show the
results from the cases with $\tau_{\rm sp}=10^{\rm 7}$ yr.
Fig.~\ref{hrdspwd} shows that the hot subdwarfs from the WD + MS
channel are divided into two groups. The first group, with
relatively higher luminosity, is from the systems for which mass
transfer begins on the MS, and the second one, with relatively
lower luminosity, is from the systems for which mass transfer
starts in the HG. The higher luminosity for the first group are
from their larger masses than the second group.

\section{BINARY POPULATION SYNTHESIS RESULTS}\label{sect:4}
Based on the parameter space for hot subdwarfs in Fig.~\ref{gsdb},
we carry out a series of BPS calculations. In this section, we
show our BPS results.

\begin{figure}
\centerline{\includegraphics[angle=270,scale=.35]{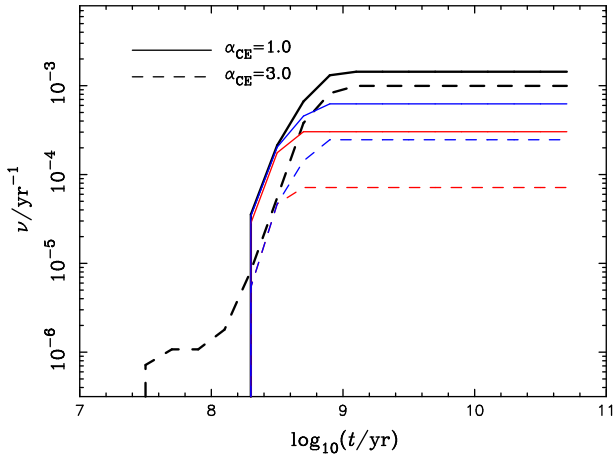}}
\caption{The evolution of the Galactic birth rates $\nu$ of the
hot subdwarf stars from the WD + MS channel for a constant star
formation rate (SFR = 5~$M_{\odot}\,{\rm yr}^{\rm -1}$) and
different $\alpha_{\rm CE}$. The blue lines show the cases with
$\tau_{\rm sp}=10^{\rm 7}$ yr, while the red lines show the cases
with $\tau_{\rm sp}=0$ yr. The black lines show the evolution of
the Galactic birth rate of SNe Ia, where the data are from
\citet{MENGXC17a,MENGXC18a}.}\label{sfr}
\end{figure}

\begin{figure}
\centerline{\includegraphics[angle=270,scale=.35]{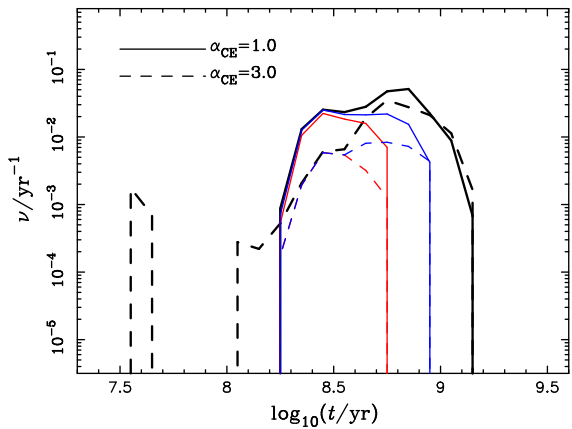}}
\caption{The evolution of the birth rates $\nu$ of the hot
subdwarf stars from the WD + MS channel for a single starburst of
$10^{\rm 11}$ $M_{\odot}$ at $t=0$ and different $\alpha_{\rm
CE}$. The blue lines show the cases with $\tau_{\rm sp}=10^{\rm
7}$ yr, while the red lines show the cases with $\tau_{\rm sp}=0$
yr. The black lines show the evolution of the birth rate of SNe
Ia, where the data are from
\citet{MENGXC17a,MENGXC18a}.}\label{single}
\end{figure}

\subsection{Birth rate}\label{sect:4.1}
Fig.~\ref{sfr} shows the Galactic birth rates of the hot subdwarfs
from the surviving companions of SNe Ia from the WD + MS channel
for different spin-down timescales and different $\alpha_{\rm
CE}$. As expected in Sec.~\ref{sect:3.3}, the birth rates from the
cases with $\tau_{\rm sp}=10^{\rm 7}$ yr are always higher than
those from the cases with $\tau_{\rm sp}=0$ yr, no matter what
$\alpha_{\rm CE}$ is. The Galactic rate of the hot subdwarfs from
the SN Ia channel is $2.3-6\times10^{\rm -4}\,{\rm yr}^{\rm -1}$
for the cases with $\tau_{\rm sp}=10^{\rm 7}$ yr and
$0.7-3\times10^{\rm -4}\,{\rm yr}^{\rm -1}$ for the cases with
$\tau_{\rm sp}=10^{\rm 7}$ yr. As expected, the birth rate of the
hot subdwarfs is significantly lower than the Galactic birth rate
of the SNe Ia from the SD channel, i.e. 7\% to 46\% of SNe Ia may
lead to hot subdwarfs, depending on different assumptions.

The evolution\footnote{Here, we neglected the time interval from
the supernova explosion to the formation of a hot subdwarf because
it is much shorter than the timescale from the formation of a
primordial binary to supernova explosion.} of the birth rate of
the hot subdwarfs from the SN Ia channel for a single starburst is
presented in Fig.~\ref{single}. Comparing the evolution with those
for SNe Ia, the evolution of the birth rate of the hot subdwarfs
has a lower peak and a shorter delay time. The delay time is
shorter than 1 Gyr, and then such hot subdwarfs belong to young
population. In addition, the hot subdwarfs from the cases with
$\tau_{\rm sp}=10^{\rm 7}$ yr have a longer delay time than those
from $\tau_{\rm sp}=0$ yr for their less massive initial
companions (see Fig.~\ref{gsdb}).

\begin{figure}
\begin{minipage}[t]{0.5\textwidth}
    \centering
\includegraphics[angle=270,scale=.33]{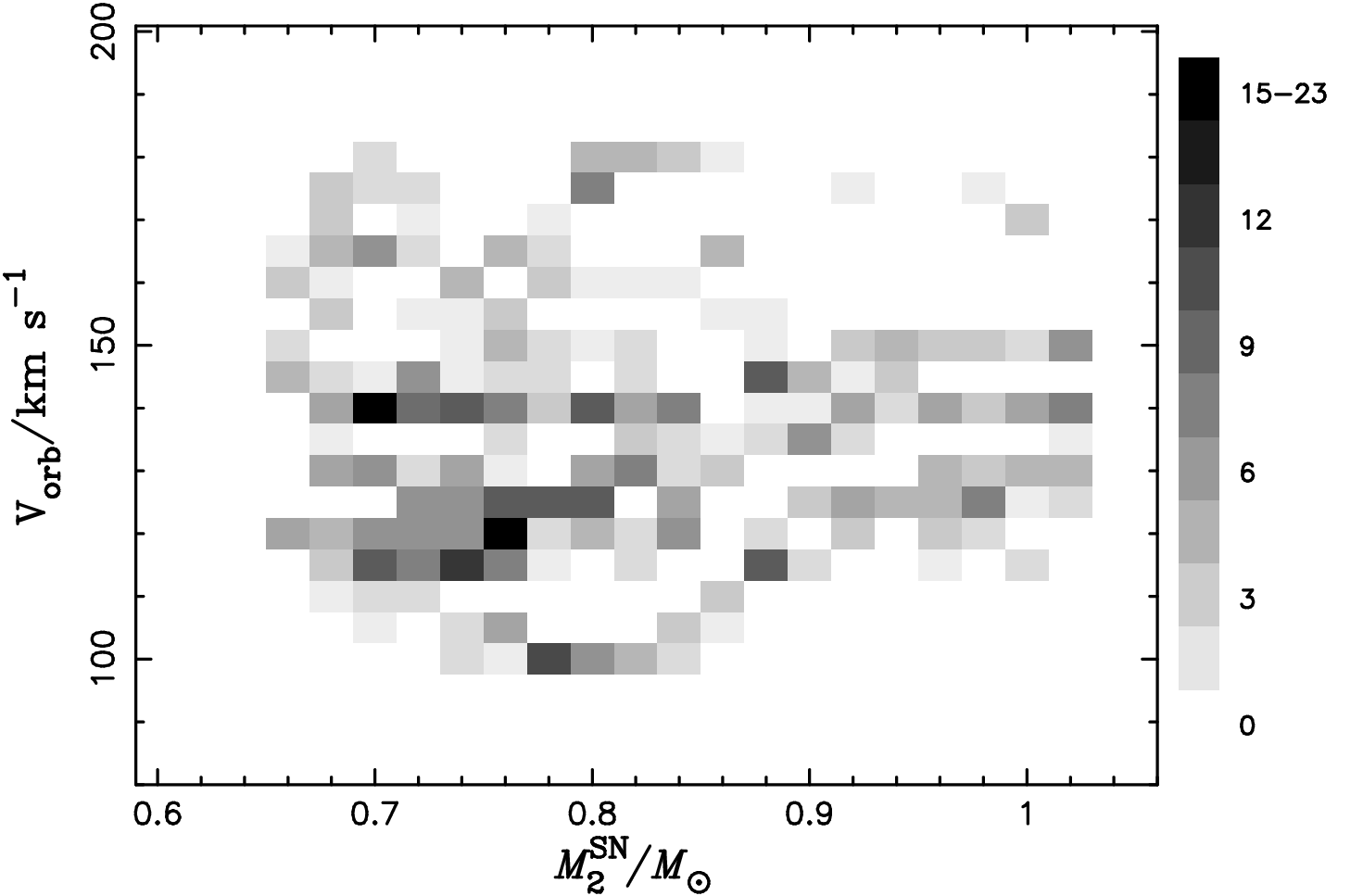}
\end{minipage}
\begin{minipage}[t]{0.5\textwidth}
    \centering
\includegraphics[angle=270,scale=.33]{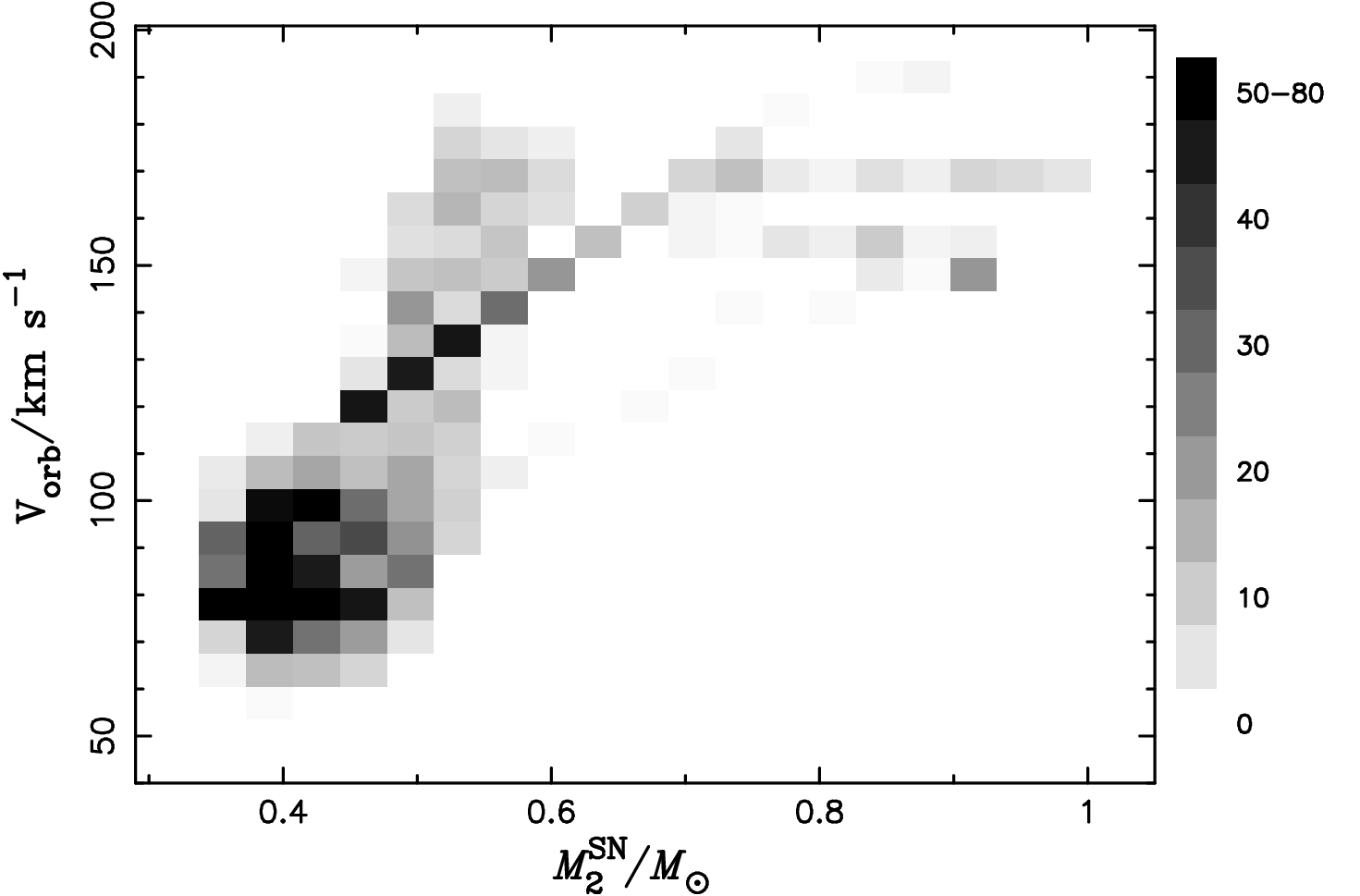}
\end{minipage}
\caption{The distribution of the companion mass and the orbital
velocity of the binary system at the moment of supernova explosion
\textbf{for} $\alpha_{\rm CE}=1.0$. Left: $\tau_{\rm sp}=0$ yr;
Right: $\tau_{\rm sp}=10^{\rm 7}$ yr.}\label{sdbvm2}
\end{figure}


\subsection{The mass and orbital velocity of the companions at supernova moment}\label{sect:4.2}
At the moment of the supernova explosion, the supernova ejecta may
impact the companion and strip off a part of its envelope and the
companion may also get a kick velocity (\citealt{MAR00};
\citealt{MENGXC07}). However, the amount of the stripped-off
material is relatively small compared to the companion mass and
the kick velocity is also relatively small compared to the orbital
velocity of the binary at the moment of the supernova explosion,
especially for the cases in which the companions are hot subdwarfs
at the supernova (\citealt{MENGXC19a}). After the supernova
explosion, we evolve the surviving companions as single stars,
where the Reimers's wind cannot significantly change its mass. So,
to great extent, the companion mass and the orbital velocity at
the moment of the supernova explosion may represent the mass and
space velocity of a surviving companion.

In Fig.~\ref{sdbvm2}, we show the distributions of the companion
mass and the orbital velocity at the moment of the supernova
explosion for the cases of $\tau_{\rm sp}=0$ yr and $\tau_{\rm
sp}=10^{\rm 7}$ yr, respectively. Here, we only show the cases
with $\alpha_{\rm CE}=1.0$ because the cases with $\alpha_{\rm
CE}=3.0$ are similar. For the case with $\tau_{\rm sp}=0$ yr, the
companion mass (orbital velocity) almost uniformly distributes
between 0.65 $M_{\odot}$ and 1.05 $M_{\odot}$ (between 100 ${\rm
km\,s^{\rm -1}}$ and 200 ${\rm km\,s^{\rm -1}}$). However, for the
case with $\tau_{\rm sp}=10^{\rm 7}$ yr, the companion mass is
between 0.35 $M_{\odot}$ and 1.0 $M_{\odot}$, but focuses on 0.4
$M_{\odot}$. The orbital velocity is between 50 ${\rm km\,s^{\rm
-1}}$ and 200 ${\rm km\,s^{\rm -1}}$, but focuses on 150 ${\rm
km\,s^{\rm -1}}$. In addition, there is a significant group in the
right panel of Fig.~\ref{sdbvm2} around $M_{\rm 2}^{\rm
SN}\simeq0.4$ $M_{\odot}$ and $V_{\rm orb}\simeq80$ ${\rm
km\,s^{\rm -1}}$, where the companions are the hot subdwarfs at
the supernova. The MS surviving companions generally have a mass
between 0.6 $M_{\odot}$ and 1.0 $M_{\odot}$ and an orbital
velocity between 150 ${\rm km\,s^{\rm -1}}$ and 200 ${\rm
km\,s^{\rm -1}}$ at the moment of the supernova explosion (see
\citealt{MENGXC19a}). Roughly, there is a correlation between the
companion mass and the orbital velocity, i.e. the higher the
companion mass, the higher the orbital velocity. This can be
easily explained by binary evolution. For a given initial WD + MS
binary system, the mass ratio of the companion to the WD may be
reversed because of the mass transfer between the WD and its
companion (\citealt{HAN04}; \citealt{MENGYANG10a}). For the case
of $\tau_{\rm sp}=10^{\rm 7}$ yr, the mass ratio for all the
systems is reversed at the supernova, and the smaller the mass
ratio (the less massive the companion), the longer the orbital
period and the smaller the orbital velocity.

\begin{figure}
\centerline{\includegraphics[angle=270,scale=.35]{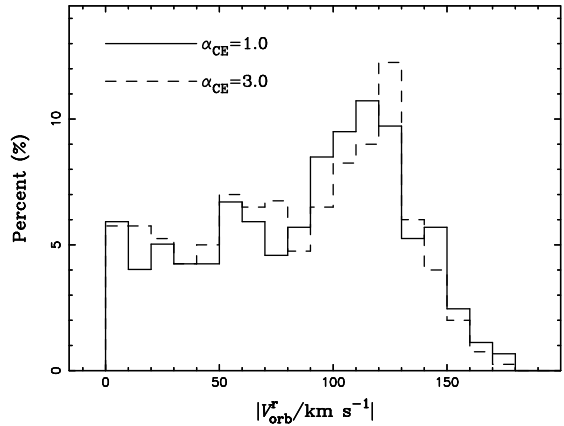}}
\caption{The distributions of the radial velocities of the hot
subdwarf surviving companions for different $\alpha_{\rm CE}$ and
$\tau_{\rm sp}=10^{\rm 7}$ yr.}\label{vorbdis}
\end{figure}

Measuring the radial velocity of stars is an important way to
search for the surviving companion of a SN Ia
(\citealt{RUIZLAPUENTE04}) and the radial velocity may be obtained
by spectral observations, such as in the LAMOST survey
(\citealt{LUOYP19}). In Fig.~\ref{vorbdis}, we show the
distributions of the radial velocities of the hot subdwarfs for
different $\alpha_{\rm CE}$, where their orbital inclinations $i$
are uniformly generated in $\cos i$. Here, we only show the cases
with $\tau_{\rm sp}=10^{\rm 7}$ yr, because a spin-down time scale
of a few $10^{\rm 6}$ yr may be necessary for SNe Ia
(\citealt{SOKER18}; \citealt{MENGXC18a}; \citealt{MENGXC18b}). The
distributions of the radial velocities for different $\alpha_{\rm
CE}$ are similar, i.e. a peak around 120 ${\rm km\,s^{\rm -1}}$
with a sharp cutoff at 180 ${\rm km\,s^{\rm -1}}$ and a long tail
to 0, because of the similar distribution of the orbital
velocities for different $\alpha_{\rm CE}$.

\begin{figure*}
\begin{center}
\epsfig{file=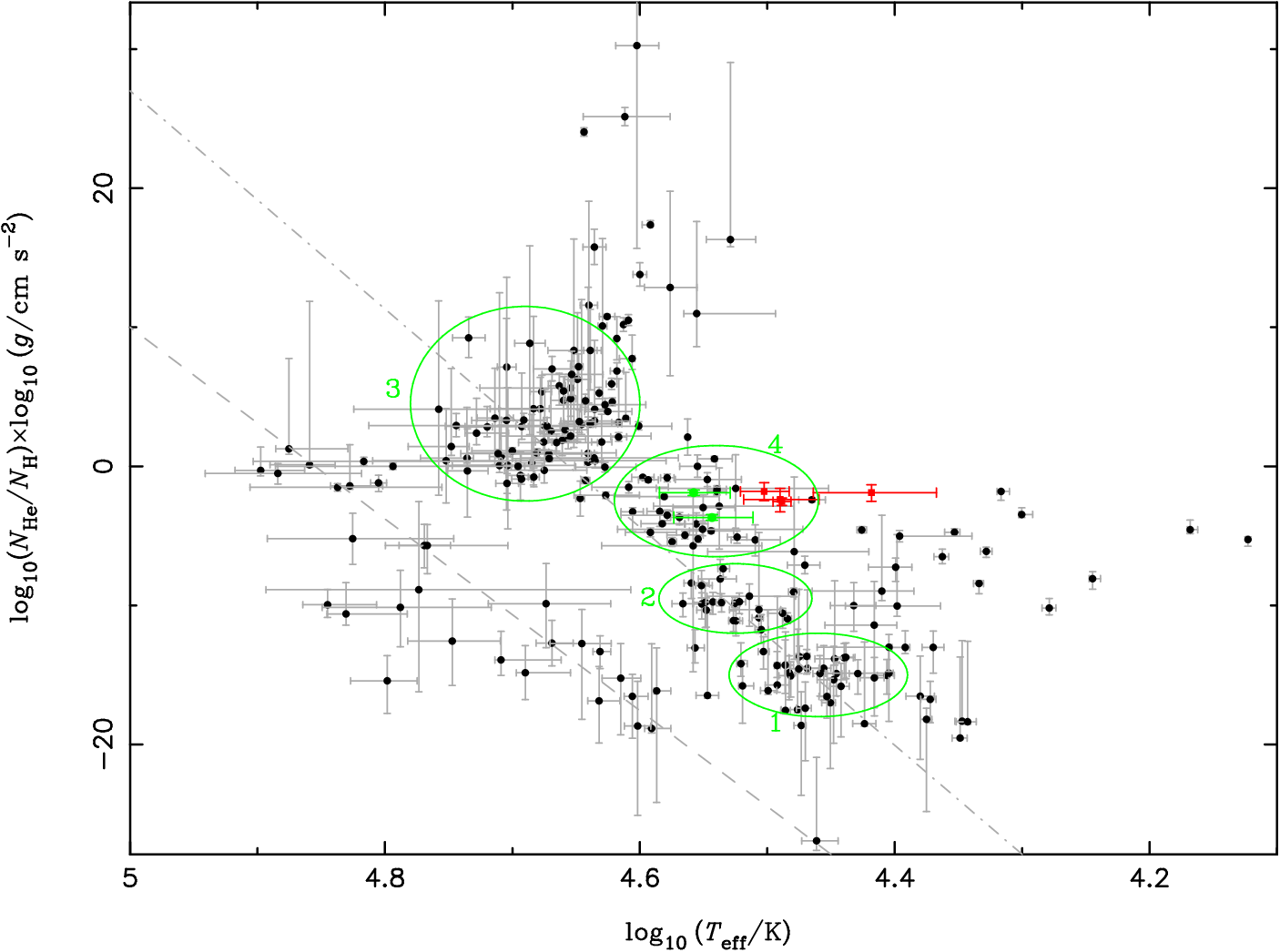,angle=270,width=12.2cm}
 \caption{$\log(N_{\rm He}/N_{\rm H})\times\log g$ vs. $\log T_{\rm
eff}$ diagram for the samples of hot subdwarf stars from the GALEX
(\citealt{NEMETH12}) and LAMOST (\citealt{LUOYP16,LUOYP19};
\citealt{LEIZX19,LEIZX20}) surveys, where the stars with
significant binary signals are excluded. Two sequences for He weak
subdwarfs and more He-rich ones are shown by dashed and dot-dashed
lines, respectively. Three groups labelled by numbers are slow
(group 1) and rapid (group 2) sdB pulsators and the hot He-rich
sdO stars (group 3), respectively (\citealt{NEMETH12}). The origin
of the group 4 named by \citet{LUOYP16} is still uncertain. Four
red squares are BLAPs from \citet{PIETRUKOWICZ17} and two green
stars are iHe-rich pulsating ones from \citet{GREEN11} and
\citet{LATOUR19}.}\label{theobv}
  \end{center}
\end{figure*}

\begin{figure}
\centerline{\includegraphics[angle=270,scale=.35]{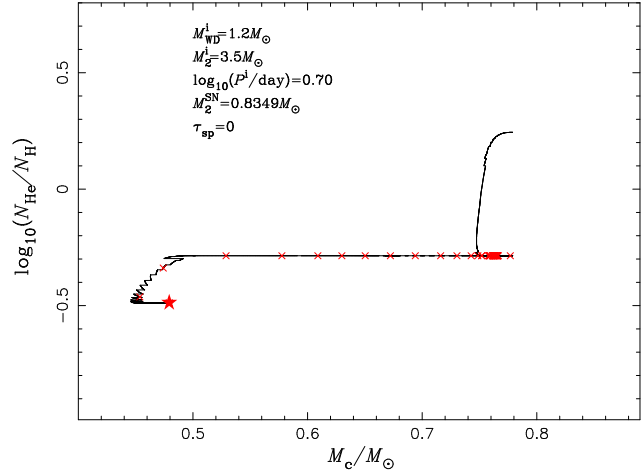}}
\caption{The evolution of the core mass and surface helium
abundance of the surviving companion from the system with the
initial parameters of [$M_{\rm WD}^{\rm i}, M_{\rm 2}^{\rm i},
\log_{\rm 10}(P^{\rm i}{\rm /day})]$=[$1.2\,M_{\odot},
3.5\,M_{\odot}, 0.7$] and $\tau_{\rm sp}=0$ yr. The red star shows
the position where the supernova explosion is assumed. The age
interval between adjacent crosses in the line is $10^{\rm 6}$
yr.}\label{xhecor}
\end{figure}

\section{COMPARISON WITH OBSERVATIONS}\label{sect:5}
\subsection{Comparison with observations}\label{sect:5.1}
Surface helium abundance is a very important factor to
differentiate different sub-classes of hot subdwarfs.
Observationally, hot subdwarfs are divided into several groups
according to their surface helium abundance and effective
temperature (\citealt{NEMETH12}). For example, about 10 to 20
percent of hot subdwarfs have a helium-rich atmosphere and such
hot subdwarfs may be further divided into two sub-classes,
extremely helium-rich stars with $\log_{\rm 10}(N_{\rm He}/N_{\rm
H})>0$ and iHe-rich stars with $-1<\log_{\rm 10}(N_{\rm He}/N_{\rm
H})<0$, where $N_{\rm He}$ and $N_{\rm H}$ are the surface number
densities of helium and hydrogen, respectively
(\citealt{LUOYP16,LUOYP19})\footnote{We notice that different
authors (\citealt{DRILLING13}; \citealt{MARTIN17};
\citealt{LUOYP16,LUOYP19}) applied different abundance
calibrations for iHe-rich stars but these calibrations only
slightly affect the relative fraction of iHe-rich stars in the
whole subdwarf population.}. The iHe-rich stars were first
discovered by \citet{NEMETH12}, but definitely confirmed to be an
independent sub-group by a sample from the survey of LAMOST, which
clearly showed a new group of hot subdwarfs with $\log_{\rm
10}(N_{\rm He}/N_{\rm H})\simeq-0.5$ and $T_{\rm eff}\simeq35000$
K in $\log (N_{\rm He}/N_{\rm H})-T_{\rm eff}$ plane
(\citealt{LUOYP16,LUOYP19}).

Here, we collected the atmospheric parameters of some hot
subdwarfs from the GALEX and LAMOST surveys (\citealt{NEMETH12},
\citealt{LUOYP16,LUOYP19}; \citealt{LEIZX19,LEIZX20}), and plotted
them into a $\log(N_{\rm He}/N_{\rm H})\times\log g-\log T_{\rm
eff}$ plane (Fig.~\ref{theobv}), where the stars with significant
binary signals are excluded. The four BLAPs
(\citealt{PIETRUKOWICZ17}) with spectral observations and two
iHe-rich pulsators (\citealt{GREEN11}; \citealt{LATOUR19}) are
also included in the figure. The figure clearly shows four groups
noticed by \citet{NEMETH12} and \citet{LUOYP16}, where group 4
represents the iHe-rich ones. Nonetheless, the figure also shows
some new features: 1) the position of the BLAPs in the
$\log(N_{\rm He}/N_{\rm H})\times\log g-\log T_{\rm eff}$ plane is
quite close to the group 4, with a slightly lower effective
temperature, showing a connection between the BLAPs and iHe-rich
hot subdwarfs; 2) there seems to exist a sub-group at the bottom
of group 3 [$\log_{\rm 10}(N_{\rm He}/N_{\rm H})\times\log_{\rm
10}(g/{\rm cm\,s^{\rm -2}})\approx0$] connecting with group 4.

By analyzing the kinematic properties of some iHe-rich hot
subdwarfs, \citet{MARTIN17} suggested that the iHe-rich stars with
halo orbits could be the surviving companion of SNe Ia.
\citet{MENGXC20} also showed that the surface helium abundance of
the hot subdwarfs from the WD + MS channel may be consistent with
the iHe-rich ones. As an example, Fig.~\ref{xhecor} shows the
evolution of the core mass and surface helium abundance of the
surviving companion from the system with initial parameters of
[$M_{\rm WD}^{\rm i}, M_{\rm 2}^{\rm i}, \log_{\rm 10}(P^{\rm
i}{\rm /day})]$=[$1.2\,M_{\odot}, 3.5\,M_{\odot}, 0.7$] and
$\tau_{\rm sp}=0$ yr after supernova explosion. The definition of
the core mass is the same as that used by \citet{HAN94} and
\citet{MENGXC08} and the difference between $M_{\rm 2}^{\rm SN}$
and $M_{\rm c}$ may roughly represent the hydrogen-rich envelope
mass of the surviving companion. After the supernova explosion,
the companion ascends the red giant branch (RGB), where its
surface helium abundance increases significantly with large-scale
convection in its envelope. After helium is ignited in the centre
of the star at the tip of the RGB, the companion becomes a
horizontal branch (HB) star and stays on the HB for a while. On
the HB, large-scale convection ceases in the envelope, and the
surface helium abundance no longer changes until the star becomes
a helium red giant, where a large-scale convective envelope
develops again, leading to a second phase of increasing surface
helium abundance. During the HB phase, shell hydrogen burning
above the helium-burning core continues to consume the
hydrogen-rich envelope. When the hydrogen-rich envelope is so thin
that the shell hydrogen burning ceases, the star becomes a hot
subdwarf with an iHe-rich atmosphere, as shown in
\citet{MENGXC20}.

However, in this paper we do not consider the effects of radiative
levitation and gravitational settling because it is still quite
uncertain in how these effects affect the surface helium abundance
of hot subdwarfs (see the discussion by \citealt{HEBER16}). If
these effects are not significant, the hot subdwarfs formed from
this channel would show the atmospheric properties of iHe-rich
ones for as long as about $10^{\rm 8}$ yr. Nonetheless, the real
time spent in iHe-rich phase is probably much shorter than that in
our models, because our models omit some of the physics. In the
following parts of the paper, we adopt the most optimistic
assumption that the hot subdwarfs from the SN Ia channel are
iHe-rich.

\begin{figure}
\centerline{\includegraphics[angle=270,scale=.35]{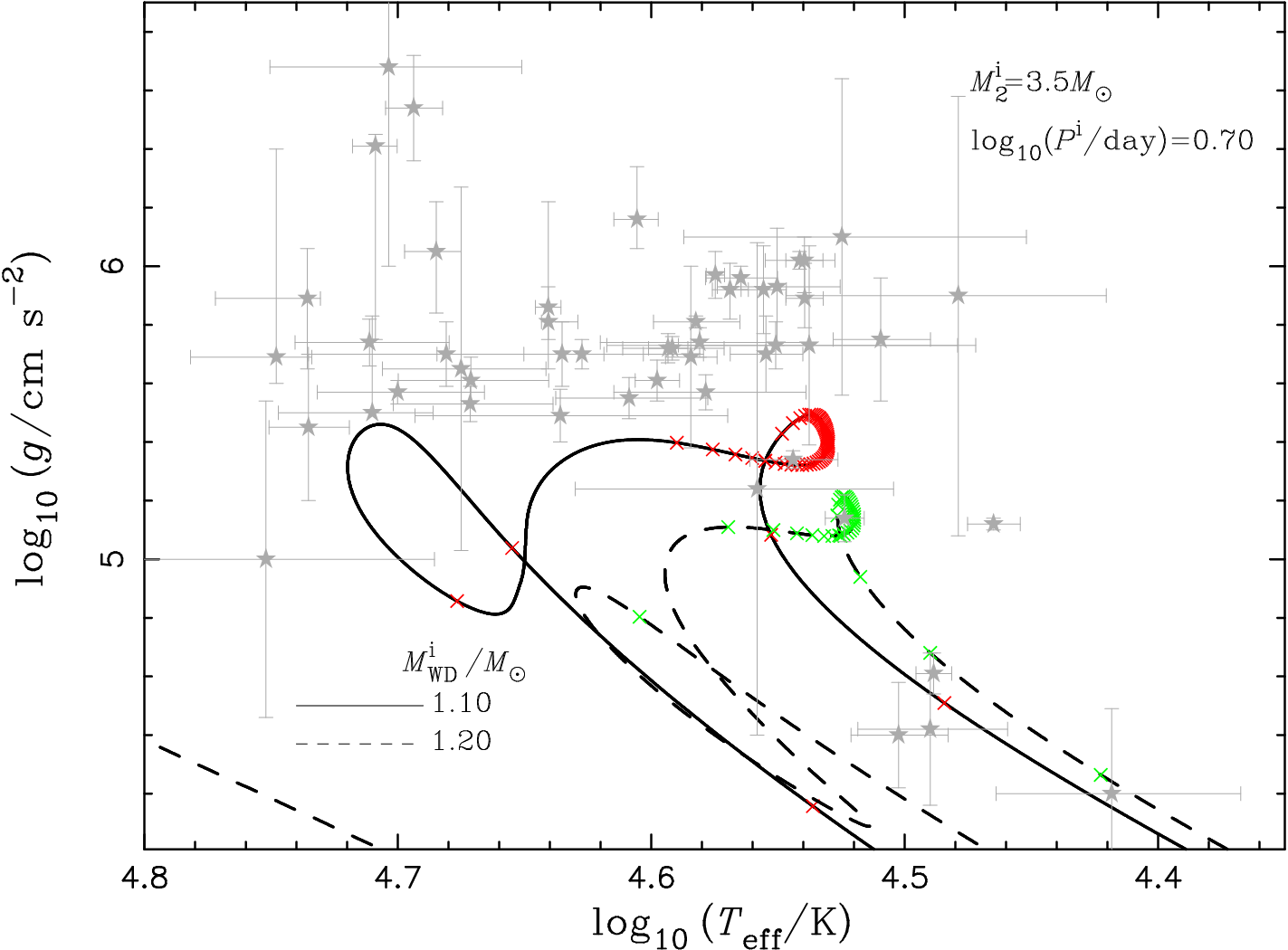}}
\caption{The evolution of the companions in $\log g-\log T_{\rm
eff}$ diagram, where the initial parameters of the binary systems
and the $\tau_{\rm sp}$ are the same to those in Fig.~\ref{hrdwd}.
The gray stars belongs to group 4 and $\log_{\rm 10}(N_{\rm
He}/N_{\rm H})\times\log_{\rm 10}(g/{\rm cm\,s^{\rm -2}})\approx0$
as shown in Fig.~\ref{theobv}. The data are from \citet{NEMETH12},
\citet{LUOYP16,LUOYP19}, \citet{LEIZX19,LEIZX20} and
\citet{PIETRUKOWICZ17}. The age interval between adjacent crosses
in every line is $10^{\rm 6}$ yr. }\label{grawd}
\end{figure}

Since we can not directly compare our results with observations in
the $\log(N_{\rm He}/N_{\rm H})\times\log g-\log T_{\rm eff}$
plane for omitting some of the physics, we want to check whether
or not the hot subdwarfs from the WD + MS channel may reproduce
the position of iHe-rich hot subdwarfs in $\log g-\log T_{\rm
eff}$ diagram. In Fig.~\ref{grawd}, we show the evolution of the
companions in a $\log g-\log T_{\rm eff}$ diagram, where the
systems are the same to those in Fig.~\ref{hrdwd}. In the figure,
we also plot some hot subdwarfs belonging to group 4 and
$\log_{\rm 10}(N_{\rm He}/N_{\rm H})\times\log_{\rm 10} (g/{\rm
cm\,s^{\rm -2}})\approx0$ as shown in Fig.~\ref{theobv}. Although
the evolutionary tracks covers some iHe-rich hot subdwarfs with
relatively lower gravities, they cannot cover most of the iHe-rich
ones. In addition, the transitional case shown in Fig.~\ref{hrdwd}
does not appear as the hot subdwarfs.

\begin{figure}
\begin{minipage}[t]{0.5\textwidth}
    \centering
\includegraphics[angle=270,scale=.33]{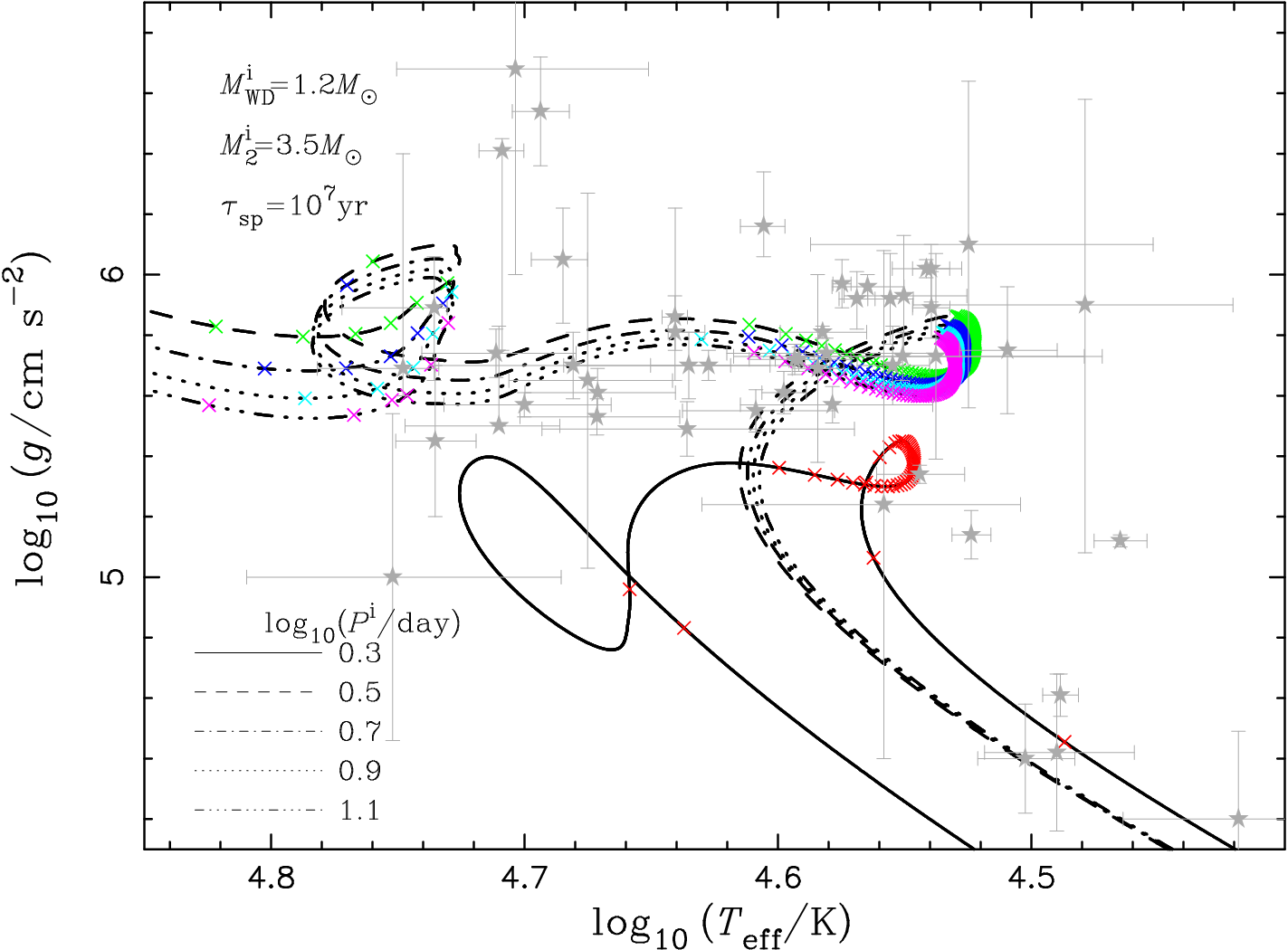}
\end{minipage}
\begin{minipage}[t]{0.5\textwidth}
    \centering
\includegraphics[angle=270,scale=.33]{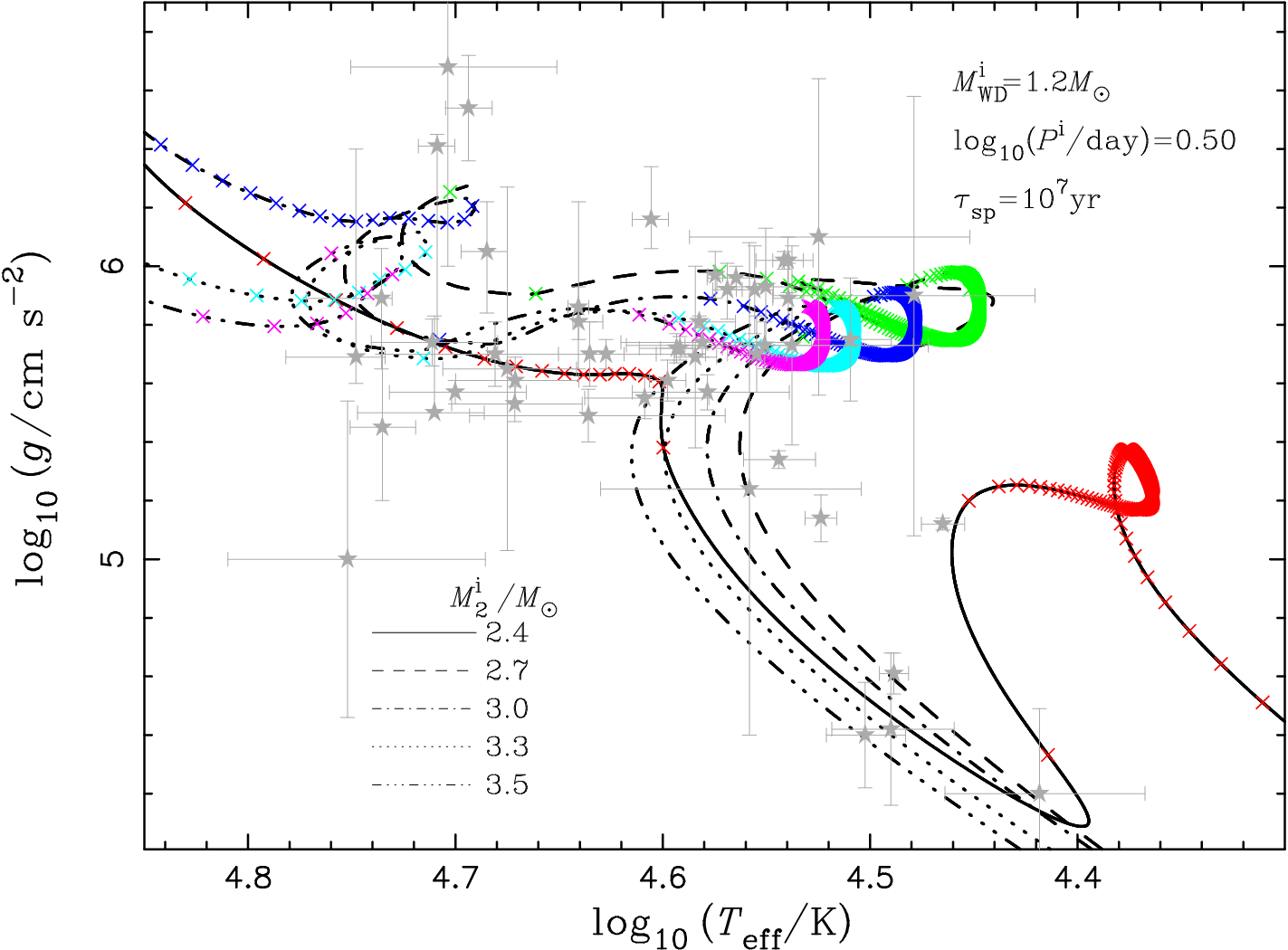}
\end{minipage}
\caption{The evolution of the companions in $\log g-\log T_{\rm
eff}$ diagram, where the initial parameters of the binary systems
and the $\tau_{\rm sp}$ are the same to those in
Fig.~\ref{hrdpersp7} (Left) and in Fig.~\ref{hrdm2sp7} (Right).
The gray stars belongs to group 4 and $\log_{\rm 10}(N_{\rm
He}/N_{\rm H})\times\log_{\rm 10} (g/{\rm cm\,s^{\rm
-2}})\approx0$ as shown in Fig.~\ref{theobv} and the data are from
\citet{NEMETH12}, \citet{LUOYP16,LUOYP19}, \citet{LEIZX19,LEIZX20}
and \citet{PIETRUKOWICZ17}. The age interval between adjacent
crosses in every line is $10^{\rm 6}$ yr. }\label{grapersp7}
\end{figure}

Similarly, in Fig.~\ref{grapersp7}, we show the evolution of the
companions in a $\log g-\log T_{\rm eff}$ diagram from systems
with $\tau_{\rm sp}\,=\,10^{\rm 7}$ yr and ($M_{\rm WD}^{\rm i}\,
,M_{\rm 2}^{\rm i}$) = ($1.2\,M_{\odot}$, $3.5\,M_{\odot}$), but
with different initial orbital periods (left panel), and with
[$M_{\rm WD}^{\rm i}, \log_{\rm 10}(P^{\rm i}{\rm /day})]$ =
[$1.2\,M_{\odot}$, 0.5] but with different initial companion
masses (right panel). The evolutionary tracks of the surviving
companions cover the iHe-rich hot subdwarfs in the $\log g-\log
T_{\rm eff}$ diagram well, i.e. the hot subdwarfs from the SN Ia
channel present the properties of group 4 stars during their
central helium burning phase, and appear as those with $\log_{\rm
10}(N_{\rm He}/N_{\rm H})\times\log_{\rm 10} (g/{\rm cm\,s^{\rm
-2}})\approx0$ when the central helium is almost exhausted. This
indicates that there could be an evolutionary connection between
the stars in group 4 and those with $\log_{\rm 10}(N_{\rm
He}/N_{\rm H})\times\log_{\rm 10} (g/{\rm cm\,s^{\rm
-2}})\approx0$.


\begin{figure}
\begin{minipage}[t]{0.5\textwidth}
    \centering
\includegraphics[angle=270,scale=.33]{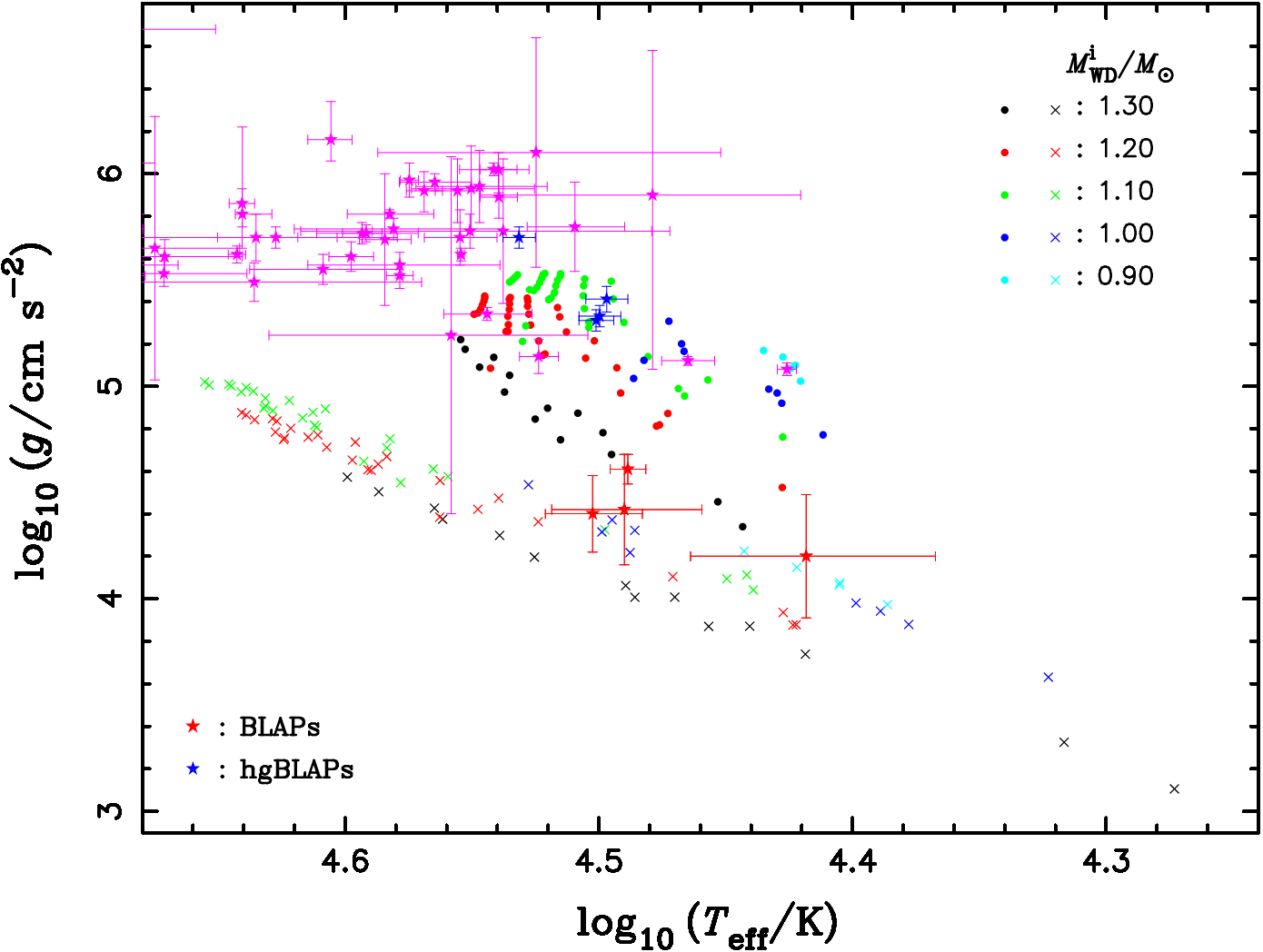}
\end{minipage}
\begin{minipage}[t]{0.5\textwidth}
    \centering
\includegraphics[angle=270,scale=.33]{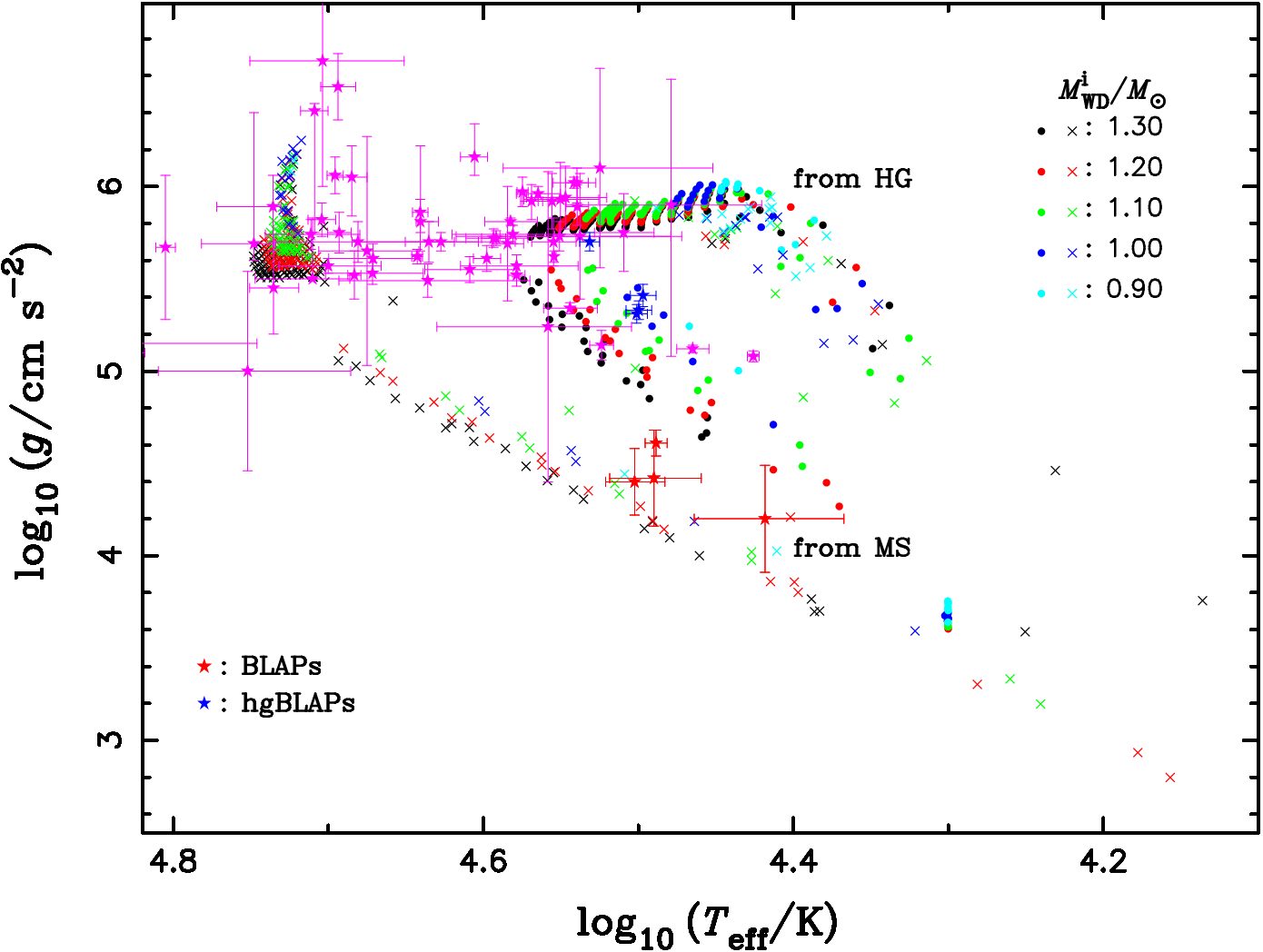}
\end{minipage}
\caption{The final outcomes of the evolution of the surviving
companions for different initial WD masses in $\log g-\log T_{\rm
eff}$ diagram. The dots represent the times that the surviving
companions during the hot subdwarf phase have the lowest
luminosity, while the crosses represent the times that the helium
is exhausted in the center of the surviving companions. The purple
stars belongs to group 4 and $\log(N_{\rm He}/N_{\rm H})\times\log
g\sim0$ as shown in Fig.~\ref{theobv} and the data are from
\citet{NEMETH12}, \citet{LUOYP16,LUOYP19} and
\citet{LEIZX19,LEIZX20}. The red stars are BLAPs from
\citet{PIETRUKOWICZ17} and the blue stars are high-gravity BLAPs
from \citet{KUPFER19}. Left: $\tau_{\rm sp}=$ yr; Right:
$\tau_{\rm sp}=10^{\rm 7}$ yr.}\label{grasp}
\end{figure}


In Fig.~\ref{grasp}, we show the outcomes of the evolution of the
surviving companions for all the models with $\tau_{\rm sp}=0$ yr
(left panel) and $\tau_{\rm sp}=10^{\rm 7}$ yr (right panel) in
the $\log g\,-\log T_{\rm eff}$ diagram, and make a comparison
with the hot subdwarfs belonging to group 4 and $\log_{\rm
10}(N_{\rm He}/N_{\rm H})\times\log_{\rm 10} (g/{\rm cm\,s^{\rm
-2}})\approx0$ as shown in Fig.~\ref{theobv}. The dots in the
figures represent the times for which the surviving companions
have the lowest luminosity during the hot subdwarf phase, while
the crosses represent the times for which helium is exhausted in
the centres. Generally, for the cases with $\tau_{\rm sp}=0$ yr,
the hot subdwarfs from the SN Ia channel have a lower gravity than
those belonging to group 4 and with $\log_{\rm 10}(N_{\rm
He}/N_{\rm H})\times\log_{\rm 10} (g/{\rm cm\,s^{\rm
-2}})\approx0$, as shown in Fig.~\ref{grawd}. However, for the
cases with $\tau_{\rm sp}=10^{\rm 7}$ yr, the outcomes may well
cover the regions of the hot subdwarfs belonging to group 4 and
those with $\log_{\rm 10}(N_{\rm He}/N_{\rm H})\times\log_{\rm 10}
(g/{\rm cm\,s^{\rm -2}})\approx0$. Again, the results here also
indicate that there could be an evolutionary connection between
group 4 and stars with $\log_{\rm 10}(N_{\rm He}/N_{\rm
H})\times\log_{\rm 10} (g/{\rm cm\,s^{\rm -2}})\approx0$.

Comparing the results in Fig~\ref{grasp}, there are two groups of
hot subdwarfs in the right panel, as shown in Fig.~\ref{hrdspwd},
while only one group in the left panel. For the group in the left
panel, the initial binary systems begin mass transfer when the
companions are in the HG (see also Fig.~\ref{gsdb}). However, in
the right panel of Fig.~\ref{grasp}, the first group, with
relatively higher surface gravity, are from the systems for which
mass transfer starts in the HG, while the second one, with
relatively lower surface gravity (i.e. the declining sequence in
the right panel), are from the systems for which transfer starts
on the MS. In particular, the group in the left panel and the
second group in the right panel cover a similar region in $\log
g-\log T_{\rm eff}$ plane. The difference between the two panels
is because that a longer spin-down timescale leads to a hot
subdwarf with a thinner hydrogen-rich envelope. The envelope
thickness of a hot subdwarf can significantly affect its
observational properties (see \citealt{XIONGHR17}). In addition,
there is a tail for the first group in the right panel. This
comprises the systems for which the supernova explosion occurs
when the companions are RG stars (see also Fig.~\ref{hrdm2sp7}).

In Fig.~\ref{grasp}, we also compare our results with the BLAPs
and high-gravity BLAPs (hgBLAPs) (\citealt{PIETRUKOWICZ17};
\citealt{KUPFER19}). Our model may reproduce the positions of the
BLAPs in the $\log g-\log T_{\rm eff}$ plane, no matter how long
the spin-down timescale. However, for the cases with $\tau_{\rm
sp}=0$ yr, the BLAPs are from the systems for which mass transfer
starts in the HG (see also \citealt{MENGXC20}), while they are
from the systems for which mass transfer starts on the MS for the
cases with $\tau_{\rm sp}=10^{\rm 7}$ yr. Based on their positions
in Fig.~\ref{grasp}, we may conclude that the BLAPs are in the
middle and late helium-burning phase, consistent with previous
studies (\citealt{WUT18}; \citealt{MENGXC20}). The hgBLAPs have
similar properties to the BLAPs except for their high surface
gravity and low surface helium abundance (\citealt{KUPFER19}). Our
model may also reproduce their positions in the $\log g\,-T_{\rm
eff}$ plane. Assuming a fundamental mode for hgBLAPs,
\citet{KUPFER19} deduced that hgBLAPs have masses ranging from
0.25 $M_{\odot}$ to 0.35 $M_{\odot}$ and then they suggested that
the hgBLAPs and the BLAPs could belong to the same class of
pulsators, composed of young helium core pre-white dwarfs.
However, if the hgBLAPs and BLAPs were the young helium core
pre-white dwarfs, there would be a substantial pulsation period
drift of $\dot{P}\simeq10^{\rm -11}$ ${\rm s}\,{\rm s}^{\rm -1}$,
which is one magnitude larger than what is observed for the BLAPs
(\citealt{PIETRUKOWICZ17}; \citealt{CALCAFERRO21}). Meanwhile, if
the pulsation modes of the hgBLAPs were the first overtone, their
masses range from 0.45 $M_{\odot}$ to 1.1 $M_{\odot}$, consistent
with the mass range of the hot subdwarfs in this paper (see
Fig.~\ref{sdbvm2}).

\subsection{The properties of the hot subdwarfs from the WD + MS channel}\label{sect:5.2}
Compared to the standard binary evolutionary origin of the hot
subdwarfs (\citealt{HANZW02,HANZW03}; \citealt{ZHANGXF17}), the
hot subdwarfs from the WD + MS channel may show some special
properties.

I) Depending on different models, the Galactic birth rate of the
hot subdwarfs from the SN Ia channel is between $7\times10^{\rm
-5}\,{\rm yr}^{\ -1}$ and $6\times10^{\rm -4}\,{\rm yr}^{\ -1}$,
which means that the SN Ia channel may contribute to 2.1 to 3.7
percent of single hot subdwarfs (\citealt{HANZW03}). The hot
subdwarfs from the channel may contribute to some iHe-rich ones,
but we can not give the exact contribution to the iHe-rich
population based on the results here, because many mechanisms are
neglected, such as gravitational settling, radiative levitation,
radiation-driven winds, surface rotation and magnetic fields, etc,
which are quite uncertain in theory (see the discussion in
\citealt{HEBER09,HEBER16}).

II) The hot subdwarfs from the SN Ia channel are single stars,
which inherit the orbital velocity of the binary systems at the
moment of the supernova explosion (\citealt{CANAL01};
\citealt{HANSEN03}; \citealt{HAN08}). Because the velocity
direction is random, some hot subdwarfs from SNe Ia could show
some special Galactic kinematical properties, such as a disk orbit
with a high eccentricity and a retrograde velocity
(\citealt{RANDALL15}; \citealt{MARTIN17}).

III) The hot subdwarfs from the SN Ia channel belong to a young
population with a delay time of less than 1 Gyr, but they do not
always associate with young populations because they may leave
their birth place due to their inherited orbital velocities. Some
surviving companions are unusual MS stars at the supernova moment
(see, e.g., the solid line in Fig.~\ref{hrdpersp7}), and may leave
their birth places by as much as tens of kpc before they become
hot subdwarfs (see also Sec.~\ref{sect:3.1.2}). The association of
a surviving hot subdwarf with a supernova remnant means that the
surviving companion were a hot subdwarf at the time of supernova
explosion, such as those of Kepler's supernova and SN 1006
(\citealt{MENGXC19a}).

IV) After the impact of the supernova ejecta on the surviving
companions, their atmosphere may be polluted by the supernova
ejecta. Therefore, some hot subdwarfs from the SN Ia channel could
show an enhancement of the iron-peak elements. However, it could
be difficult to distinguish the enhancement from that derived from
radiative levitation.

\subsection{Constraints from Galactic kinematical observations}\label{sect:5.3}
The hot subdwarfs from the SN Ia channel may show some special
Galactic kinematical properties (\citealt{HANSEN03} and
\citealt{JUSTHAM09}). \citet{MARTIN17} studied the kinematics of
88 hot subdwarfs and suggested that the iHe-rich ones with extreme
halo orbits could originate from the surviving companions of SNe
Ia, which reveals a clue on the contribution of the SN Ia channel
to the iHe-rich hot subdwarfs, i.e. 2 or 3 in 27 iHe-rich hot
subdwarfs are possible from SN Ia channel. In this section, we
replicate a similar analysis to \citet{MARTIN17} by using a large
sample of 747 hot subdwarfs from the LAMOST survey in
\citet{LUOYP19} to constrain the potential contribution of the hot
subdwarfs from the SN Ia channel to the iHe-rich one. The volume
completeness of the sample is more than 92\% (\citealt{LUOYP20}).
\citet{LUOYP19} divided the sample into four subclasses based on
the surface helium abundance $y=N_{\rm He}/N_{\rm H}$, where
$N_{\rm He}$ and $N_{\rm H}$ are the number densities of helium
and hydrogen, respectively. Based on the classifications by
\citet{LUOYP19}, we named the groups with $\log_{\rm 10} y\geq0$
as He-rich, $0>\log_{\rm 10} y\geq-1$ as iHe-rich, $-1>\log_{\rm
10} y\geq-2.2$ as He-deficient and $\log_{\rm 10} y<-2.2$ as
extremely helium deficient (eHe-deficient). In the sample, the
completeness of the He-deficient group is relatively better than
the others and the He-deficient group has the standard binary
evolutionary origin (\citealt{LUOYP19}). So, we here only compare
the Galactic kinematical properties of the iHe-rich group with the
He-deficient group.

If the SN Ia channel partly contribute to the iHe-rich hot
subdwarfs, due to the random direction of the inherited orbital
velocity, it is expected that the iHe-rich hot subdwarfs have a
similar average space velocity to that of the He-deficient ones
but with a larger standard deviation, and the fraction of the
iHe-rich stars with a high/low space velocity would be larger than
that of the He-deficient group. For the same reason, the
distributions of the Galactic kinematical properties of the
iHe-rich hot subdwarfs, e.g. the orbital eccentricity, the orbital
maximum vertical amplitude ($z_{\rm max}$), $z$-component of the
orbital angular momentum ($J_{\rm z}$), the apocentre radius
($R_{\rm ap}$) and the pericentre radius ($R_{\rm peri}$) of the
orbit in the Galaxy, would be different from those of the
He-deficient ones.

\begin{figure}
\begin{minipage}[t]{0.5\textwidth}
    \centering
\includegraphics[angle=270,scale=.32]{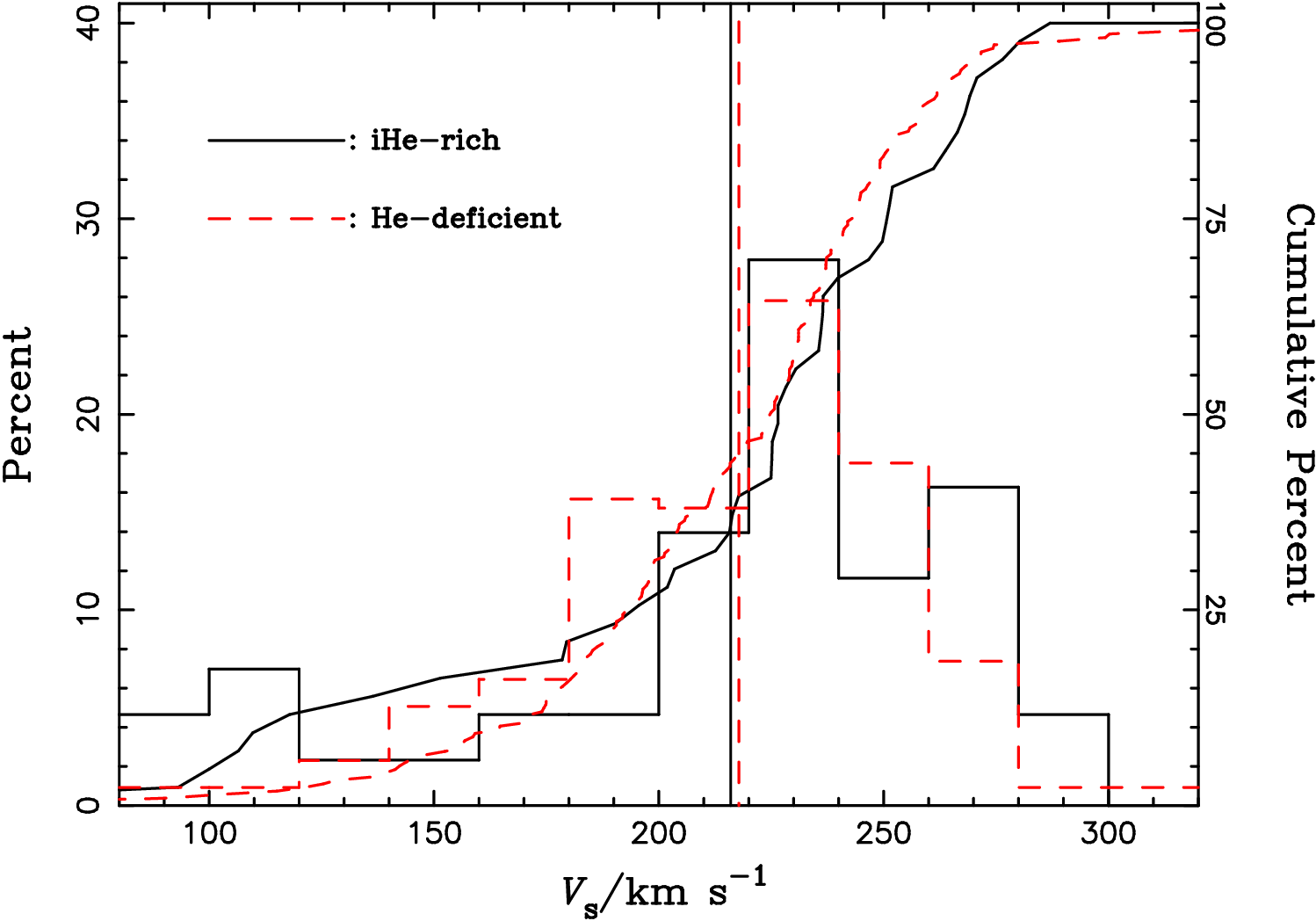}
\end{minipage}
\begin{minipage}[t]{0.5\textwidth}
    \centering
\includegraphics[angle=270,scale=.32]{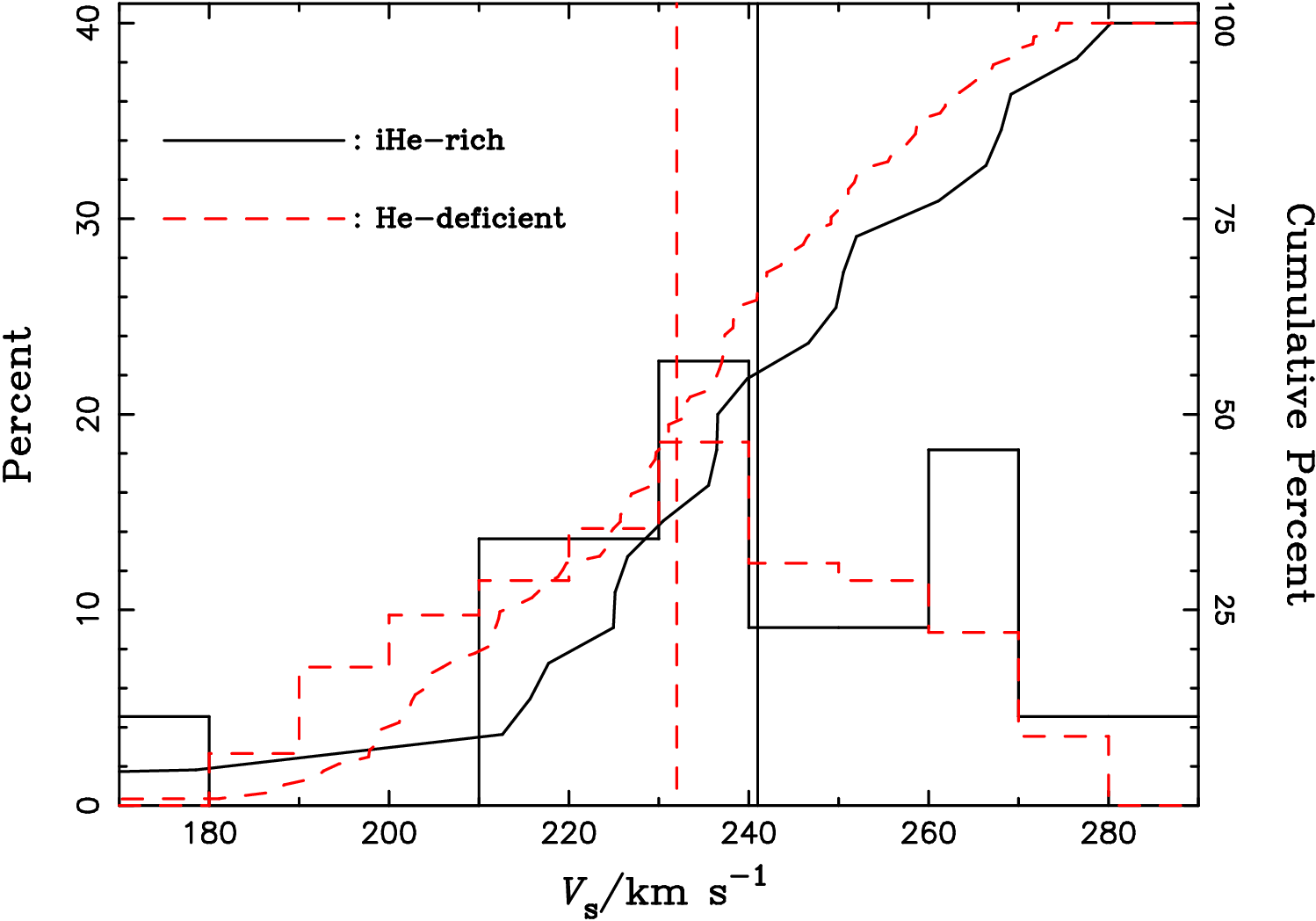}
\end{minipage}
\caption{The distributions and the cumulative distributions of the
space velocities for the iHe-rich (black solid lines) and the
He-deficient (red dashed lines) groups, respectively. The vertical
lines show the mean values of the distributions. Left: the whole
sample; Right: the thin disk sample. The data are from
\citet{LUOYP19}. }\label{vtdis2}
\end{figure}

Actually, both \citet{MARTIN17} and \citet{LUOYP19} noticed that
iHe-rich hot subdwarfs show a more diverse kinematic distribution
than He-deficient ones, and suggested that the higher dispersion
are from the pollution by halo population. In Fig.~\ref{vtdis2},
we show the distributions of the space velocities of the
He-deficient and iHe-rich stars, where $V_{\rm s}=(U^{\rm
2}+V^{\rm 2}+W^{\rm 2})^{\rm 1/2}$, and $U$, $V$ and $W$ are the
Cartesian Galactic velocities, directed toward the Galactic
centre, the Galactic rotation and North Galactic Pole,
respectively. The left panel is for the total sample and the right
one is for the thin disk sample. As expected, the distribution of
the space velocities between the He-deficient and the iHe-rich
groups are similar, e.g. a similar mean space velocity. The
Kolmogorov--Smirnov (K--S) tests give a $p$-value of 0.578 and
0.396 for the total and thin disk samples, respectively, where
$p$-value represents the probability that two distributions are
chosen from the same underlying population. In addition, the
distributions for the iHe-rich group show a higher standard
deviation, as noticed by \citet{MARTIN17} and \citet{LUOYP19}. In
particular, the iHe-rich group have a higher fraction of high/low
space velocities than those for the He-deficient sample and the
fraction of the iHe-rich stars with a high space velocity for the
thin disk sample is even higher than that from the sample with the
halo population (see Table\,\ref{Tab:1}). So, it seems difficult
to explain the differences of the distributions of the space
velocities between the He-deficient and the iHe-rich stars solely
by a pollution mechanism from halo stars, while the inherited
orbital velocity from the binary system at the supernova moment
could partly contribute to the difference.

\begin{table*}
  \caption{Statistical results of the distributions of the space velocity for iHe-rich and He-deficient subsamples,
  i.e. number of the sample (Column 2), mean space velocity (Column 3), standard deviation
 (Column 4), fractions of $V_{\rm s}>260\,{\rm km\,s^{\rm -1}}$
(Column 5) and $V_{\rm s}<180\,{\rm km\,s^{\rm -1}}$ (Column 6).
The `total' means the total subsample and the `thin' means the
thin disk populations. }\label{Tab:1}
  \begin{center}
    \begin{tabular}{lccccc}
      \hline
subsample  & number & $\overline{V}_{\rm s}$  & $\sigma$
 & fraction ($V>260\,{\rm km\,s^{\rm -1}}$) & fraction ($V<\,180{\rm km\,s^{\rm -1}}$)\\
  &  & (${\rm km\,s^{\rm -1}}$) & (${\rm km\,s^{\rm -1}}$)
 & (\%) & (\%) \\
      \hline
        iHe-rich (total)      & 43  & 216 & 52 & 20.9 & 20.9\\
        He-deficient (total)  & 216 & 218 & 44 & 10.1 & 15.7\\
        iHe-rich (thin)       & 22  & 241 & 34 & 27.3 & 4.5\\
        He-deficient (thin)   & 122 & 232 & 27 & 12.3 & 0.0\\
      \hline
    \end{tabular}
  \end{center}
\end{table*}

\begin{figure}
\centerline{\includegraphics[angle=270,scale=.35]{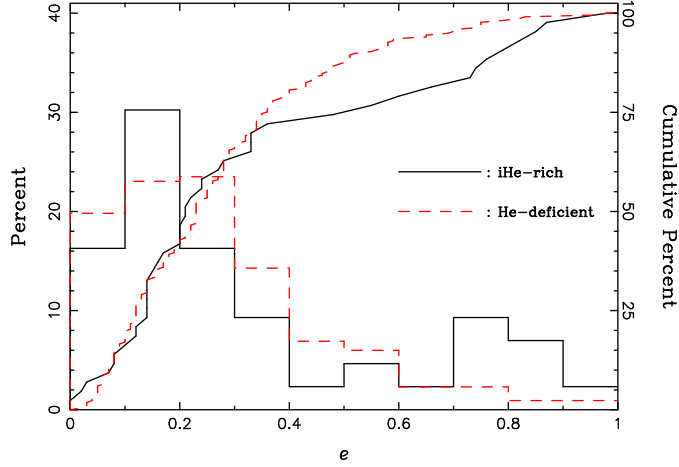}}
\caption{The distribution and the cumulative distribution of the
orbital eccentricity in the Galaxy for He-deficient (red dashed
lines) and iHe-rich (black solid lines) groups, respectively. The
data are from \citet{LUOYP19}.}\label{eccen}
\end{figure}

In Fig.~\ref{eccen}, we show the distributions of the orbital
eccentricity in the Galaxy for the He-deficient and iHe-rich
groups, respectively. The two distributions are similar and a K--S
test gives a $p$-value of 0.239. Similarly, the distributions of
$z_{\rm max}$,  $J_{\rm z}$, and $R_{\rm ap}$ and $R_{\rm peri}$
of the orbit in the Galaxy look similar. For simplicity, we do not
show the distributions. 2D K--S tests for $z_{\rm max}$ and
$J_{\rm z}$ and for $R_{\rm ap}$ and $R_{\rm peri}$ gives a
$p$-value of 0.359 and 0.307, respectively. So, the distributions
of the Galactic kinematical properties between the iHe-rich and
He-deficient subgroup are indistinguishable, which indicate that
the SN Ia channel cannot be the dominant contributor to the
iHe-rich hot subdwarfs.


At present, we still cannot give a definitive conclusion on how
large the contribution from the SN Ia channel to the iHe-rich
popoulation is . \citet{MENGXC20} suggested that BLAPs are
helium-core-burning stars from SN Ia channel and live in the
iHe-rich phase for a few $10^{\rm 7}$ yr
(\citealt{PIETRUKOWICZ17}). If the hot subdwarfs from the SN Ia
channel also stay in the iHe-rich phase for a few $10^{\rm 7}$ yr,
the channel could contribute to the whole population of iHe-rich
hot subdwarfs by a few percent. This is roughly consistent with
the fraction difference between the iHe-rich stars and the
He-deficient ones with high/low space velocities, and also
consistent with the hint in \citet{MARTIN17}, i.e. 2 or 3
candidates in 27 iHe-rich hot subdwarfs are from SN Ia channel.

\section{DISCUSSIONS}\label{sect:6}
In this paper, following \citet{MENGXC20}, we study the hot
subdwarfs from the surviving companions of SNe Ia in details, and
give the birth rate of the hot subdwarfs from the WD + MS channel.
Such hot subdwarfs may present some properties of the iHe-rich
ones. In this section, we discuss the uncertainties of this work
and the other possible origin of the iHe-rich subdwarfs.

\begin{figure}
\centerline{\includegraphics[angle=270,scale=.35]{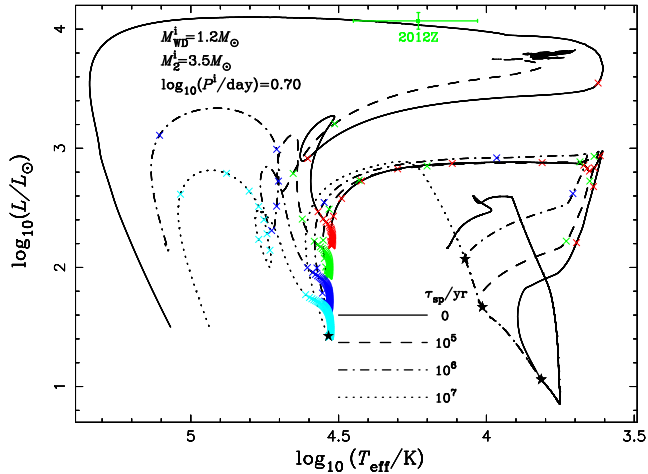}}
\caption{The evolution of the companions in the HR diagram for
different assumed spin-down timescales, $\tau_{\rm sp}$, where the
initial parameters of the system are [$M_{\rm WD}^{\rm i}, M_{\rm
2}^{\rm i}, \log_{\rm 10}(P^{\rm i}{\rm /day})]$=[$1.2\,M_{\odot},
3.5\,M_{\odot}, 0.9$]. The black stars show the positions where
supernova explosions are assumed. The age interval between
adjacent crosses in every line is $10^{\rm 6}$ yr. The green cross
with label `2012Z' presents the suggested companion of SN 2012Z
(\citealt{MCCULLY14}).}\label{hrdsp}
\end{figure}

\subsection{Uncertainties}\label{sect:6.1}
In this paper, we only choose two limiting cases for spin-down
timescale. Here, we choose a system with initial parameters of
[$M_{\rm WD}^{\rm i}, M_{\rm 2}^{\rm i}, \log_{\rm 10}(P^{\rm
i}{\rm /d})$]=[$1.2\,M_{\odot}, 3.5\,M_{\odot}$, 0.7] as an
example to check the effect of $\tau_{\rm sp}$ on the following
evolution of the surviving companion in the HR diagram, as shown
in Fig.~\ref{hrdsp}. For this binary, the donor star fills its
Roche lobe in the HG, and then the WD accretes hydrogen-rich
material from the donor increasing its mass. After about
$2\times10^{\rm 5}$ yr, the WD mass reaches $M_{\rm
WD}=1.378~M_{\odot}$ at which mass it would explode
(\citealt{NTY84}). Then, the WD enters into the pseudo spin-down
phase, where $\tau_{\rm sp}$ is set to be $0$ yr, $10^{\rm 5}$ yr,
$10^{\rm 6}$ yr and $10^{\rm 7}$ yr. $1.04\times10^{\rm 5}$ yr
after $M_{\rm WD}=1.378~M_{\odot}$, the WD stops mass-increasing
phase although the mass-transfer rate is still as high as
$3.5\times10^{\rm -7}\,M_{\odot}\,{\rm yr^{\rm -1}}$, at which
[$M_{\rm WD}, M_{\rm 2}, \log_{\rm 10}(P{\rm
/d})$]=[$1.5992\,M_{\odot}, 0.7162\,M_{\odot}$, 0.1350]. Then,
$2.4\times10^{\rm 6}$ yr after $M_{\rm WD}=1.378~M_{\odot}$, the
mass transfer ceases. This case clearly shows that the exact time
of the onset of the spin-down phase is quite unclear and depends
on different assumptions (see the discussions in
\citealt{MENGXC13} in details).

In Fig.~\ref{hrdsp}, no matter how long is $\tau_{\rm sp}$, the
surviving companion from this initial binary system may evolve to
a hot subdwarf. However, a longer $\tau_{\rm sp}$ leads to a lower
luminosity of the hot subdwarf. This is because a longer
$\tau_{\rm sp}$ means a longer mass-loss phase and then a less
massive surviving companion. The masses of the surviving
companions for $\tau_{\rm sp}=0$, $10^{\rm 5}$, $10^{\rm 6}$ and
$10^{\rm 7}$ yr are 0.8349, 0.7182, 0.6100 and 0.5342 $M_{\odot}$,
respectively. So, for a given initial binary system, different
$\tau_{\rm sp}$ will mainly affect the mass and then the
luminosity of the hot subdwarf.

At the moment of supernova explosion, the supernova ejecta may
impact on the companion and strip off some of their envelope, and
at the same time, the companion may get a kick velocity
(\citealt{MAR00}; \citealt{MENGXC07}). We neglected the
stripped-off effect and the kick velocity in this paper because
they are not significant. If the interaction between supernova
ejecta and the companion is considered, the upper limit to the
masses (space velocities) of the surviving companions is slightly
lower (higher) than those shown in Fig.~\ref{sdbvm2}. In addition,
the stripping can slightly change the initial parameter space for
the hot subdwarfs, and then the Galactic birth rates of the hot
subdwarfs from the SN Ia channel is slightly higher than those
shown in Fig.~\ref{sfr}.

In this paper, we have not considered the WD + RG channel. After a
supernova explosion, almost all the envelope of the companion from
the WD + RG channel has been stripped off and the surviving
companion consists of a helium core and a thin hydrogen-rich
envelope (\citealt{MAR00}). If the core {mass of the surviving
companion at the moment of the supernova explosion is lower than
the lower limit to ignite helium, the surviving companion becomes
a single low-mass WD (\citealt{JUSTHAM09}; \citealt{MENGYANG10b}).
Nevertheless, some surviving companions from the WD + RG channel
may have helium cores as massive as 0.45 $M_{\odot}$, and then
they may become hot subdwarfs after ignition of helium
(\citealt{JUSTHAM11}; \citealt{MENGXC13}). So, the WD + RG channel
may also produce single hot subdwarfs. However, detailed BPS
results always give a quite low rate of SNe Ia from the WD + RG
channel, although some arguments exist for the exact birth rate
from the WD + RG channel (\citealt{HAC99b}; \citealt{HAN04};
\citealt{CHENXC11}).

Although the hot subdwarfs from the WD + MS channel may reproduce
some properties of iHe-rich ones and then the WD + MS channel
could contribute some iHe-rich subdwarfs, we can not give the
exact contribution from the SN Ia channel to the iHe-rich
population because many mechanisms are neglected here, such as
gravitational settling, radiative levitation, radiation-driven
winds, surface rotation and magnetic fields, etc. The life time of
a hot subdwarf star at the iHe-rich phase is heavily dependent on
how these mechanisms play any kind of role, but this is quite
uncertain in theory (see the discussion in
\citealt{HEBER09,HEBER16}). For example, some iHe-rich hot
subdwarf stars show a strong enrichment of heavy elements, but
some more helium-rich ones do not. This indicates that radiative
levitation plays a key role in forming their atmospheric chemical
pattern, but radiative levitation could be affected by the helium
abundance (\citealt{DORSCH19}; \citealt{NASLIM20}).

\subsection{The other possible channel contributing to the iHe-rich hot subdwarf stars}\label{sect:6.2}
Although there are still some puzzles with the formation of hot
subdwarfs (\citealt{HEBER09,HEBER16}), it is believed that binary
interaction via one or two common-envelope (CE) ejection phases
plays a key role to their formation (\citealt{HANZW02,HANZW03};
\citealt{HEBER09}). The merging of two helium white dwarfs which
experience two common-envelope ejection phases may well explain
the properties and the formation of the extremely helium-rich
subdwarfs (\citealt{HANZW03}; \citealt{ZHANGXF12}). However, it is
still unclear about the formation of iHe-rich ones, although
several evolutionary scenarios are suggested, such as the late
hot-flasher scenario (\citealt{DCRUZ96}; \citealt{MOEHLER04};
\citealt{MILLER08}), a post-common-envelope system in which the
stratification of its atmosphere is incomplete
(\citealt{NASLIM12}) and the merging of a He WD with a low-mass
main-sequence star (\citealt{ZHANGXF17}).

As discussed in Section. \ref{sect:5.3}, the SN Ia scenario may
only contribute to a small fraction of the iHe-rich hot subdwarfs
and other channels or mechanisms would contribute the special hot
subdwarfs. For example, at least one iHe-rich subdwarf,
CD-20$^{\circ}$1123, is in a close binary (\citealt{NASLIM12};
\citealt{MARTIN17}), and this means that the standard binary
evolution channels contribute to a few iHe-rich hot subdwarfs.
\citet{ZHANGXF17} suggested a merging channel of a He WD and a low
mass MS star to explain the formation of iHe-rich hot subdwarfs.
According to the BPS results of \citet{ZHANGXF17}, the Galactic
birth rate of the iHe-rich hot subdwarfs from the merging channel
of a helium WD + low-mass MS star is $7.57\times10^{\rm -4}\,{\rm
yr}^{\ -1}$ and the Galactic birth rate of the hot subdwarfs from
the merging channel of two helium WDs is about $3.7\times10^{\rm
-3}\,{\rm yr}^{\ -1}$. Because the typical lifetime of a hot
subdwarf is $10^{\rm 8}$ yr and the typical lifetime of the
iHe-rich hot subdwarfs from the merging channel of a helium WD +
low-mass MS star is $2.5\times10^{\rm 6}$ yr (see Fig. 5 in
\citealt{ZHANGXF17}), the typical number ratio of the iHe-rich to
the He-rich ones is $(7.57\times10^{\rm
-4}\times2.5\times10^{\rm6})/(3.7\times10^{\rm
-3}\times10^{\rm8})\simeq5\times10^{\rm -3}$, which is too low to
compare with the observed ratio (0.347, \citealt{LUOYP19}). The
late hot-flasher scenario could also produce the iHe-rich hot
subdwarfs, but a detailed comparison between the iHe-rich hot
subdwarfs in $\omega$ Cen with the late-flasher evolutionary
tracks shows that the late-flasher scenario has difficulty to
explain the position of the iHe-rich hot subdwarfs in $\omega$ Cen
in the $\log g$ -- $T_{\rm eff}$ plane (\citealt{LATOUR14};
\citealt{HEBER16}). \citet{NASLIM11} suggested that the iHe-rich
hot subdwarfs could be from merging double WDs as He-rich ones,
and then their special atmosphere represents a snapshot of an
evolving surface chemistry (see \citealt{NASLIM12,NASLIM13}). This
suggestion gets support from some observational facts and some
theoretical hints. First, in the sample of \citet{LUOYP19}, about
22 percent of the hot subdwarfs have a helium abundance of
$\log_{\rm 10} y>-1$, consistent with the BPS results that 11 to
24 percent of the hot subdwarf stars are from merged double WDs
(\citealt{HANZW03}). Second, both He-rich and iHe-rich samples
have similar Galactic kinematics (\citealt{MARTIN17};
\citealt{LUOYP19}). More evidence comes from observation of the
globular cluster $\omega$ Cen, in which there is a remarkable
group significantly separated from other hot subdwarfs in the
$\log y-T_{\rm eff}$ plane, with helium abundances of
$-1.3<\log_{\rm 10} y<1$, and no obvious gap exists between the
He-rich and iHe-rich hot subdwarfs (see Fig. 23 of
\citealt{HEBER16}). In addition, considering that $\omega$ Cen is
a globular cluster or the core of a dwarf galaxy, its high
fraction of the hot subdwarfs from merged double WDs is also
consistent with theoretical expectations (\citealt{HAN08b}).
Moreover, the merging channel could produce rapid rotating stars
and the rapid rotation could play a kind of role for the special
surface chemistry of the hot subdwarfs in $\omega$ Cen
(\citealt{TAILO15}).

\begin{figure}
\centerline{\includegraphics[angle=270,scale=.40]{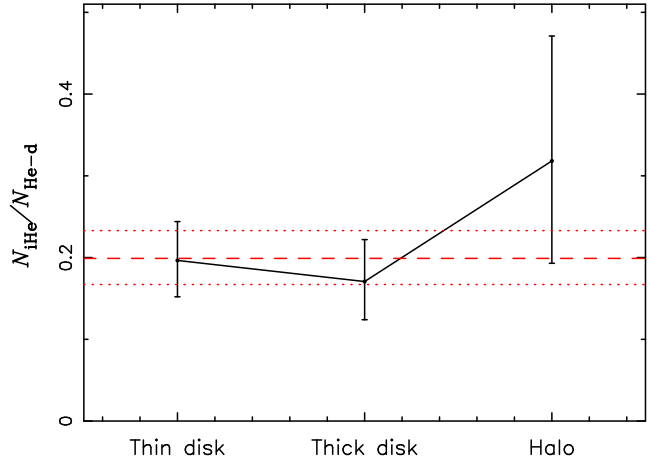}}
\caption{The number ratio of the iHe-rich to He-deficient hot
subdwarfs in the thin disk, thick disk and halo, where the error
bars are calculated assuming a binomial distribution
(\citealt{CAMERON11}). The red dashed line shows the average value
of the number ratio and the two dotted lines represent the error
range of the average value, based on the binomial distribution.
The data are from Table 6 in \citet{LUOYP19}.}\label{number}
\end{figure}

\citet{MARTIN17} argued that if the atmospheres of the iHe-rich
hot subdwarfs represent a snapshot of an evolving surface
chemistry, the iHe-rich subdwarfs should share the kinematical
properties of the helium-deficient ones, as we shown in
Sec.~\ref{sect:5.3}. In addition, \citet{MARTIN17} suggested that
analyzing the population of the iHe-rich hot subdwarfs may provide
constraint on their origin. However, for the relatively small size
of their sample, \citet{MARTIN17} did not get a statistically
significant conclusion. The hot subdwarf sample in \citet{LUOYP19}
is much larger than that in \citet{MARTIN17}. We try to make a
similar analysis to \citet{MARTIN17} by this sample. Roughly, we
may assume a constant star formation rate and a single star burst
to represent the star formation history of the Galactic disk and
halo, respectively. So, if most of the iHe-rich ones were to
represent a snapshot of an evolving surface chemistry of merged
double WDs, the number ratio of the iHe-rich to He-deficient hot
subdwarf stars in the Galactic halo would be larger than that in
the thin disk and/or the thick disk (\citealt{HANZW03};
\citealt{HAN08b}; \citealt{LUOYP19}). In Fig.~\ref{number}, we
show the number ratio of the iHe-rich to He-deficient hot
subdwarfs in the thin disk, the thick disk and the halo. The
number ratio in the thin disk and the thick disk is similar,
consistent with the assumption of a constant star formation rate
in the Galactic disk, but the number ratio in the halo is larger
than those in the thin and thick disks. However, the error bar of
the number ratio in the halo is still large because of the
statistics ($7/22$)\footnote{If the sample in \citet{LUOYP20} is
combined, the number ratio of the iHe-rich to He-deficient hot
subdwarfs in the halo is $N_{\rm iHe}/N_{\rm
He-d}=0.367^{+0.143}_{-0.118}$, which is significantly higher than
those in the thin and thick disks, but with a smaller error bar.}.

\begin{figure}
\centerline{\includegraphics[angle=270,scale=.40]{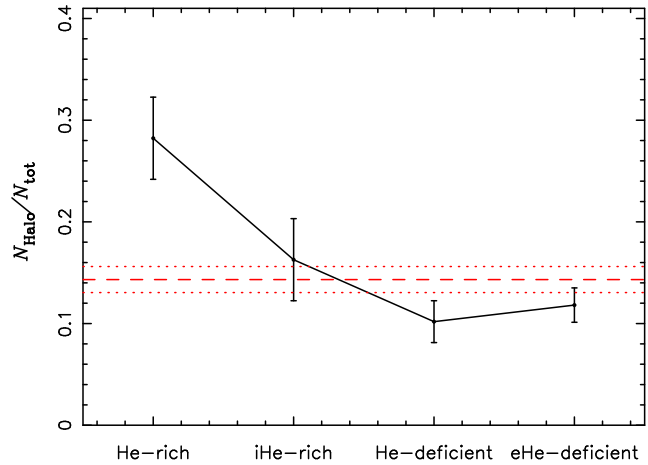}}
\caption{The number fraction of halo stars in He-rich, iHe-rich,
He-deficient and eHe-deficient subsamples, respectively, where the
error bars are calculated assuming a binomial distribution
(\citealt{CAMERON11}). The red dashed line shows the average value
of the number fraction and the two dotted lines represent the
error range of the average value, based on the binomial
distribution. The data are from Table 6 in
\citet{LUOYP19}.}\label{number2}
\end{figure}

Moreover, to some extent, the fraction of halo stars in each
subsample may represent the age of the subsample, because the
larger the number ratio, the older the subsample
(\citealt{MARTIN17}). Because the iHe-rich hot subdwarfs from the
SN Ia channel belong to a relatively young population, if majority
of the iHe-rich ones are from the merging channel of two He WDs,
the age of the iHe-rich hot subdwarfs could be younger than the
He-rich ones, but older than the He-deficient ones. Therefore, it
is expected that the fraction of halo stars in the iHe-rich
subsample is smaller than in the He-rich one, but larger than in
the He-deficient and eHe-deficient ones. Fig.~\ref{number2} shows
the fraction of halo stars in different subsamples, where the
fraction for the He-rich subsample is the largest. As expected,
the fraction for the iHe-rich subsample is lower than that for the
He-rich one, but significantly larger than those for the
He-deficient and eHe-deficient ones\footnote{The fraction of
eHe-deficient subdwarfs seems slightly larger than that of
He-deficient ones. This is mainly due to a selection effect in the
sample of \citet{LUOYP19}, which overestimates the fraction of
halo stars in the eHe-deficient group (see the detailed discussion
of \citealt{LUOYP19}).}. So, the results in Figs~\ref{number} and
~\ref{number2} are consistent with the expectation that the
iHe-rich ones are mainly from the merged double WDs and SN Ia
channel contributes some of them.

\section{CONCLUSIONS}\label{sect:7}
Some of the surviving companions of SNe Ia from the WD + MS
channel may evolve to hot subdwarfs. In this paper, we preform
detailed stellar evolutionary calculation of the surviving
companions from close WD + MS systems with different initial
parameters for the spin-up/spin-down model and the canonical
non-rotating model, and obtain the initial parameter space in
which the surviving companion can evolve to hot subdwarfs. The
parameter space for the spin-up/spin-down model is larger than
that for the canonical non-rotating model. Based on the parameter
space, we carry out a series of BPS calculation, and obtain the
Galactic birth rate of the hot subdwarfs from the SN Ia channel,
which is $2.3-6\times10^{\rm -4}\,{\rm yr}^{\rm -1}$ for the
spin-up/spin-down model and $0.7-3\times10^{\rm -4}\,{\rm yr}^{\rm
-1}$ for the canonical non-rotating model. We also show the
distribution of some integral properties of the hot subdwarfs,
e.g. the mass, the space velocity and the radial velocity. The hot
subdwarfs from the SN Ia channel have a delay time of less than 1
Gyr, and then belong to relatively young population. However, some
such hot subdwarfs may leave their birth place by as far as tens
of kpc because of an inherited orbital velocity. The hot subdwarfs
from the WD + MS channel can reproduce some observational
properties of the iHe-rich hot subdwarfs, e.g. $\log g$ and
$T_{\rm eff}$, especially for our spin-up/spin-down model. So, the
WD + MS channel is possible to contribute some iHe-rich hot
subdwarfs. We also discuss the possible origins of the iHe-rich
hot subdwarfs. An analysis of the population of a large LAMOST
sample in \citet{LUOYP19} is consistent with the previous
suggestion that most of the iHe-rich hot subdwarfs are more
possible from merged double WDs.

\section*{ACKNOWLEDGMENTS}
We are grateful to both the onymous referee, Christopher Tout, the
anonymous referee, and Zhengwei Liu for their detailed comments
and suggestions which help us to improve the manuscript greatly.
This work was supported by the NSFC (Nos. 11973080, 11733008 and
U173111). We acknowledge the science research grants from the
China Manned Space Project with NO. CMS-CSST-2021-B07. X.M.
acknowledges the support by the Yunnan Ten Thousand Talents Plan -
Young \& Elite Talents Project, and CAS `Light of West China'
Program.

\section*{DATA AVAILABILITY}
No new data were generated in support of this research, and all
the data used in this article will be shared on request to the
corresponding author.

\label{lastpage}
\end{document}